\documentclass[12pt, draftclsnofoot, onecolumn]{IEEEtran}
\ifCLASSOPTIONcompsoc
  \usepackage[nocompress]{cite}
\else
  \usepackage{cite}
\fi
\ifCLASSINFOpdf
\else
\fi
\usepackage{amssymb}
\usepackage{amsthm}
\usepackage{amsmath}
\usepackage{algorithm}
\usepackage{algpseudocode}
\usepackage{array}
\ifCLASSOPTIONcompsoc
  \usepackage[caption=false,font=footnotesize,labelfont=sf,textfont=sf]{subfig}
\else
  \usepackage[caption=false,font=footnotesize]{subfig}
\fi
\usepackage{url}
\usepackage{multicol}
\usepackage{mathtools, nccmath}
\usepackage{graphicx}
\usepackage{enumerate}
\usepackage{enumitem}
\usepackage{xcolor}

\usepackage{hyperref}
\hypersetup{
    colorlinks = true
}

\DeclareMathOperator*{\argmax}{arg\,max}

\newcommand{\myparagraph}[1]{\vspace{2mm} \noindent \textbf{\textsf{#1}}}

\newcommand{\pe}[1]{\langle #1 \rangle}
\newcommand{\se}[1]{\{ #1 \}}

\newcommand{\MX}{\infty}

\newcommand{\KK}[1]{}
\renewcommand{\KK}[1]{{\textbf{\color{blue} [@@Kai:{#1}] }}}

\newtheorem{lemma}{Lemma}
\newtheorem{definition}[lemma]{Definition}

\newtheorem{proposition}[lemma]{Proposition}
\newtheorem{assumption}{Assumption}

\usepackage[margin=1mm, skip=2pt]{caption}
\begin{document}
%{UAV Swarm Cooperation in Path Planning and Servicing Dynamic Demands}
\title{Multi-UAV Cooperative Trajectory for Servicing Dynamic Demands and Charging Battery}

\author{
Kai~Wang, Xiao~Zhang,~\IEEEmembership{Member,~IEEE}, Lingjie~Duan~\IEEEmembership{Senior Member,~IEEE} and~Jun~Tie%
\IEEEcompsocitemizethanks{
\IEEEcompsocthanksitem K. Wang, X. Zhang and J. Tie are with the College of Computer Science, South-Central University for Nationalities, Wuhan, China. (E-mails: kai.wang@my.cityu.edu.hk, xiao.zhang@my.cityu.edu.hk, tiejun@mail.scuec.edu.cn).
K. Wang is also with Department of Information Engineering, The Chinese University of Hong Kong, Shatin, NT, Hong Kong SAR, China.
L. Duan is with Engineering Systems and Design, Singapore University of Technology and Design, Singapore 487372. (E-mail: lingjie\_duan@sutd.edu.sg).
\IEEEcompsocthanksitem
A preliminary version of this paper appeared in the Proceedings of IEEE Global Communications Conference 2020 \cite{wang20uav}.}%
}

\IEEEtitleabstractindextext{%
\begin{abstract}
Unmanned Aerial Vehicle (UAV) technology is a promising solution for providing high-quality mobile services (e.g., edge computing, fast Internet connection, and local caching) to ground users, where a UAV with limited service coverage travels among multiple geographical user locations (e.g., hotspots) for servicing their demands locally. How to dynamically determine a UAV swarm's cooperative path planning to best meet many users' spatio-temporally distributed demands is an important question but is unaddressed in the literature. To our best knowledge, this paper is the first to design and analyze cooperative path planning algorithms of a large UAV swarm for optimally servicing many spatial locations, where ground users' demands are released dynamically in the long time horizon. Regarding a single UAV's path planning design, we manage to substantially simplify the traditional dynamic program and propose an optimal algorithm of low computation complexity, which is only polynomial with respect to both the numbers of spatial locations and user demands. After coordinating a large number $K$ of UAVs, this simplified dynamic optimization problem becomes intractable and we alternatively present a fast iterative cooperation algorithm with provable approximation ratio $1-(1-\frac{1}{K})^{K}$ in the worst case, which is proved to obviously outperform the traditional approach of partitioning UAVs to serve different location clusters separately. To relax UAVs' battery capacity limit for sustainable service provisioning, we further allow UAVs to travel to charging stations in the mean time and thus jointly design UAVs' path planning over users' locations and charging stations. Despite of the problem difficulty, for the optimal solution, we successfully transform the problem to an integer linear program by creating novel directed acyclic graph of the UAV-state transition diagram, and propose an iterative algorithm with constant approximation ratio. Finally, we validate the theoretical results by extensive simulations.
\end{abstract}

}

\maketitle

\IEEEdisplaynontitleabstractindextext

% \fi
\IEEEpeerreviewmaketitle

%\IEEEraisesectionheading{\section{Introduction}\label{sec:introduction}}

\IEEEraisesectionheading{\section{Introduction}\label{sec:introduction}}
\IEEEPARstart{U}{AV}
technology recently emerges as a promising solution to provide high-quality mobile services (e.g., edge computing, high-speed Internet access, and local caching) to ground users (e.g., \cite{wang20uav, asheralieva2019hierarchical, TGCN2021}). Unlike the conventional wireless communication systems, UAVs offer line-of-sight (LoS) wireless links with users \cite{saad2019vision}, which greatly improves the quality of service. Besides, due to their agility and mobility, UAVs can be deployed fast to achieve seamless wireless coverage and provide on-demand mobile services to the users under emergency conditions \cite{zhang2018fast, LiangTMC2020}. Despite of these advantages, a UAV has small service coverage due to limited antenna size and low transmit power, and needs to fly closely to service ground users \cite{wang2018traffic, xu2018uav, jiangchun2021proactive}. To service many dynamic users across different spatial locations sustainably, it is necessary for UAVs to intelligently cooperate with each other by taking into account their locations, trajectories, battery charging and the users' demands over time, and how to determine UAVs' cooperative path planning is an important question in many applications.

\subsection{Related works}\label{sec:related}
In the literature of wireless communications and networking, there are some studies of UAV deployment schemes to service ground users (e.g., \cite{pan2012cooperative, ProcIEEE2019}), given their distribution in a geographical area.
Zeng and Zhang \cite{zeng2017energy} consider a UAV-enabled base station to service multiple users for achieving maximum throughput per user, by jointly optimizing the transmit power and the UAV trajectory.
Without knowing users' spatial distribution, Xu et al. \cite{xu2018uav} study the UAV-user interaction for learning users' real locations before the single UAV's deployment.
To meet users' instantaneous traffic requests over different locations, Wang et al. \cite{wang2018traffic} study how a single UAV should dynamically adapt its location to user movements following predefined random processes.
Wang and Duan \cite{wang2018dynamic}\cite{zhewangTMC} further consider the energy allocation of a UAV for meeting dynamically arriving users' demands.
Yang et al. \cite{yang2019energy} studied energy efficient UAV path planning with energy harvesting.

{Jiang} and {Swindlehurst} \cite{Heading2012} present a heading algorithm to position an UAV for improving uplink communications for ground users. Xu et al. \cite{NFZ2020} consider various uncertainties (e.g., user location, wind speed, and no-fly zones) and jointly optimize an UAV's path planning and transmit beamforming vector for the overall energy consumption, which is a non-convex optimization problem.
In \cite{PowerTransfer2018}, UAV's trajectory is optimally designed in an UAV-enable wireless transfer system in order to maximize the total energy transfer. The cooperative trajectories of UAV-swarm are rarely exploited to improve the system performance, e.g., energy efficiency and service quality.
%The UAV with low delay is introduced in \cite{saad2019vision} and the energy efficient routing for UAV was studied in \cite{yang2019energy}.

%
In term of drone/vehicle routing in the broader literature, there are some studies on multiple vehicular data mules' joint path planning (\cite{sugihara2011path, zhang2015data}).  %somasundara2004mobile
%When the demand information are unknown in advance, Enright et al. studied the scheduling for UAVs in a dynamic environment \cite{enright2015uav}, in which targets (user demands) arrive at random and remain active until some UAV reaches the target’s location.
Zhang et al. \cite{zhang2015data} studied data sensing and transmission problem with energy consideration and proposed a two-stage data gathering approach for dynamic sensing and routing. As to path planning of UAVs, Sun et al. \cite{sun2019optimal} studied 3D trajectory design for UAV communication systems and  considered ground users' static demands only. These works do not explicitly study the UAV-to-UAV cooperation to meet users' dynamic demands spatially and temporally, and there is a lack of studies of such UAV-swarm cooperation for dynamic path planning.

%There are a few studies on UAV-swarm planning by using artificial intelligence tools %\cite{6198334, xu2020optimized}. In \cite{6198334}, the authors apply the genetic %algorithm and the particle swarm optimization algorithm to search UAV's optimal %trajectories in a complex 3D environment. In \cite{xu2020optimized}, Cheng et al. adopt %swarm intelligence algorithm to generate UAVs' path planning such that each UAV can reach %the mission area quickly and reduce the probability of being captured and destroyed. %However, such techniques suffer from high computational resources and tends to be trapped %in local optimum.

%In such works, travelling salesman problem (TSP) is formulated for each data mule to travel through multiple sensor nodes and collect sensor data once and for all, without looking at the dynamic demands over the time domain.
%However in general, no demand can be serviced in the worst case scenario and one can imagine the case where a new demand always appears at the location when no UAV is nearby and disappears when some UAV arrives the location.

%The solutions from previous works act like one-time deployment of a UAV where the UAV visits each location for at most once. The problem becomes difficult if the UAV can visit some locations for multiple times and there is a lack of research for routing UAVs repeatly over spatial locations.

We are aware that in the literature of artificial intelligence and robotics, there are some preliminary designs of cooperative vehicle trajectories to visit target locations once and for all (\cite{6198334, xu2020optimized, TNSE2021}).
% ([8] [9] [10] [11] ACTUALLY CAN REDUCE TO ONLY 2 OR 3 REFERENCES ).
%6198334, xu2020optimized, sugihara2011path, Zhao2005infocom
However, in practice the UAV-swarm may need to repeatedly visit the same set of user locations (e.g., shopping malls and other hotspots) over time according to users' activity patterns and waiting time deadline, and there is another temporal domain to optimize other than the spatial domain.
As the sustainable UAV servicing over many users is also limited by the UAV's energy capacity (\cite{zhang2020energy} \cite{yao2019qos}), we should allow UAVs to return to a charging station to recharge or change their batteries.
These problems are not addressed in previous studies by designing UAVs' cooperative path planning over users' locations and charging stations for achieving sustainable servicing.

\subsection{Main contributions }\label{sec:contr}
Given the aforementioned review, no prior optimization technique can be applied to this challenging problem, and we are interested in developing new algorithms for cooperative UAV-swarm path planning with provable performance bounds. When there are many spatial locations or user demands to service over time, our problem becomes complex and it is challenging to design  tractable algorithms with low computational complexity.

Our key novelty and main contributions in this paper are summarized as follows.

%[label={\large\textbullet}]

\begin{itemize}
\item \emph{Dynamic UAV-swarm cooperation to service users' spatial-temporal demands (Section~\ref{sec:problem-description}):} To our best knowledge, this is the first paper to design and analyze cooperative mutli-UAV trajectory planning algorithms for dynamically servicing many spatial demand locations over time. Unlike existing routing problems (e.g.,\cite{sugihara2011path, 6198334, xu2020optimized}), we consider the practical and challenging problem that the UAV-swarm need to repeatedly visit users' locations and charging stations  over a long time, and design low-complexity algorithms for guiding UAV-swarm cooperation with provable performance guarantee.
\item \emph{Optimal path planning for a single UAV (Section~\ref{sec:single-drone}):}
Regarding a single UAV's dynamic trajectory planning problem, we manage to substantially simplify the dynamic programming problem and propose a fast algorithm for returning the UAV's optimal path planning. The algorithm's complexity is low and regardless of the scale of time domain, only polynomial with respect to both the numbers of user locations and user demands.
\item \emph{Cooperative path planning of a large UAV-swarm to service many user demands (Section~\ref{sec:multiple-drone}):} When a large number $|K|$ of UAVs are cooperating to service many user locations, the simplified dynamic optimization problem becomes intractable and we alternatively present a fast iterative cooperation algorithm with provable approximation ratio $1-(1-\frac{1}{|K|})^{|K|}$ in the worst case, which arbitrarily approaches to constant guarantee $1-1/e$. Our approximation algorithm is proved to obviously outperform the traditional approach of partitioning UAVs to serve different user/location clusters separately.
\item \emph{Refined algorithm design for UAV-swarm by adding UAV charging stations} (Section~\ref{sec:energy}):
For achieving sustainable service provisioning, we jointly design UAVs' cooperative path planning over spatial-temporal users' demands and various charging stations. The problem becomes more challenging and we successfully transform it to an integer linear programming by creating novel directed acyclic graph (DAG) of the UAV-state transition diagram. To further lower the complexity we accordingly propose an iterative algorithm with constant approximation ratio.
\end{itemize}

\section{System Model and Problem Description }\label{sec:problem-description}

\begin{figure}[!t]%[htbp]
\centerline{
\includegraphics[width = 0.95 \columnwidth]{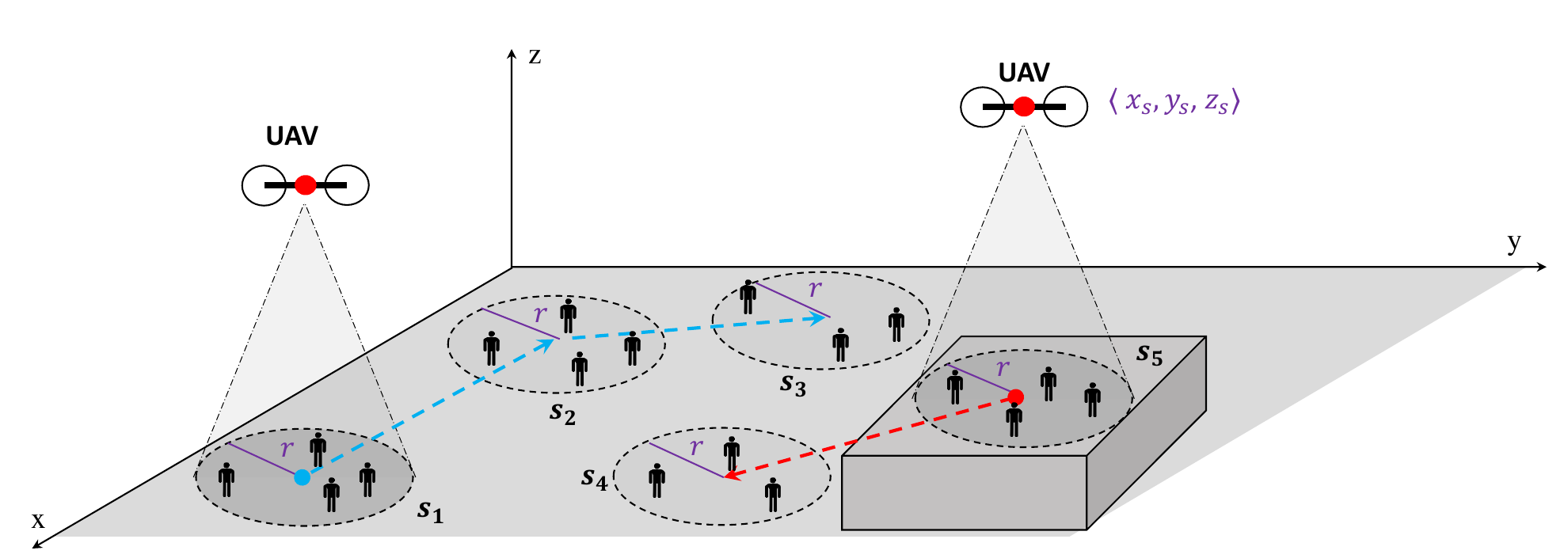}
}
\caption{An illustration of the system model, where two UAVs cooperatively route and service $|S|=5$ locations or clusters with radii $r$ in the ($x, y ,z$) 3D space. Each location (e.g., location $s_1$ here) releases a series of user demands over time and each user demand requests to be serviced by a UAV within user's waiting time. The connected blue dashed and red dashed arrows show the first and second UAVs' trajectories, respectively.}
\label{fig:model}
\end{figure}

\iffalse
\begin{figure}[!t]%[htbp]
\centerline{
\includegraphics[width = 0.8 \columnwidth]{images/drone_model.pdf}
}
\caption{An illustration of the system model, where two UAVs cooperatively route and service $|S|=5$ locations on the 2D ground plane. Each location (e.g., location 1 here) releases a series of user demands in set $J_s$ with $s=1$ over time and each demand $j \in J_s$ requests to be serviced by a UAV within time window $[r_j, d_j)$. The connected blue (black) arrows show the first (second) UAV's routing path.}
\label{fig:model}
\end{figure}
\fi

As illustrated in Fig.~\ref{fig:model},
given a set $K$ of UAVs to service a set $J$ of dynamically arriving users over a set of potential locations (e.g., shopping malls and other hotspots) across the 3D space, we seek to design UAV-swarm's cooperative path planning among these locations to meet delay-sensitive users' demands as many as possible.
%
%We have a set $K$ of UAVs to provide delay-sensitive service to dynamically arriving  users in a set $S$ of potential locations (e.g., shopping malls and other hotspots). As illustrated in Fig.~\ref{fig:model}, the UAV-swarm aims to meet as many user demands as possible, by routing UAVs among different locations.
% we consider a set $S$ of spatial locations (i.e., hotpots such as shopping malls and parks) where
%
To successfully service a user demand, a UAV must reach the corresponding location  before the waiting deadline of the demand.
The objective is to determine cooperative path planning for the UAV-swarm so that the total number of successfully serviced demands is maximized.
In the following, we detail the system model for the dynamic demands, UAV service, and summarize the key notations in Table~\ref{tbl:notation}.

As in \cite{zeng2017energy,ProcIEEE2019}, we adopt the air-to-ground model where the UAVs are deployed to the positions above ground and the wireless communication channels between UAVs and users are dominated by LoS links. LoS links are expected for air-to-ground channels in many scenarios. Therefore, the channel power gain from the UAV to each user $k$ is modeled as the free-space path loss model, i.e., $g_k = \xi \bar{d_k}^{-2}$, where $\xi$ denotes the channel power gain at a reference distance. $\bar{d_k}$ is the link distance between the UAV $i$ and ground user $k$. Given a standard transmission power $P$, the signal-to-noise ration (SNR) at ground user $k$ is given by $\gamma_k = \frac{P g_k}{\sigma^2}$, where $\sigma^2$ denotes the noise power at each ground user. We say a ground user $k$ at any location can be serviced by a UAV if the SNR at user $k$ is no less than a threshold value $\gamma_{th}$, and the data rate is $\log (1+\gamma_{th})$. Thus, we can obtain each UAV's wireless coverage disk on the ground with range $r$ and flying altitude $H$ to serve users nearby to meet target SNR as follows.
\begin{equation}\label{eq:radii}
r = \sqrt{\frac{P \xi}{\gamma_{th} \sigma^2} - H^2}.
\end{equation}

Given the spatial distribution of the ground users and the coverage radii $r$, the ground users can be clustered into the location set $S$ by applying the classical clustering algorithm \cite{hochbaum1985best}.
As illustrated in Fig.~\ref{fig:model}, each cluster of users is a circular region with ground radius $r$,
correspondingly, the associated UAV location in 3D is $s = \langle x_s,y_s,z_s \rangle \in S$, thus we denote $J_{s}$ as the set of user demands in this cluster, which are dynamically released over time.
%
%within coverage radius $r$ of UAV $i$, the associated UAV location $s_i = \langle x_i,y_i,z_i \rangle \in S$, a set $J_{s_i}$ of user demands are dynamically released over time.
Our following algorithms are thus designed based on such wireless model and set $S$.

We assume that the UAVs have sufficient bandwidth resources so that all UAVs can be assigned orthogonal channels and for achieving interference-free \cite{8660516}. Note that in our designed algorithms, multiple UAVs servicing the same UAV location simultaneously is prohibited. In practice, the assigned channels for distant UAVs can be reused during servicing. Thus, the interference among UAVs can be ignored, and henceforth we focus our study on the UAV-swarm path planning.

%released \rv{within radius $r$} of

Each user demand $j \in J_s$ associated by UAV location $s \in S$ is characterized by its release time $r_j$ and waiting deadline $d_j$, where the time difference $d_j - r_j$ tells the user's maximum waiting time.
%\footnote{Without much loss of generality, we assume $r_j$ and $d_j$ are integers. In practice, one can discrete their values for engineering purpose.}.
For each demand, it takes a fixed amount of time $q$ to service this demand.
We denote $J = \cup_{s \in S} J_s$ as the set of demands at all the locations and denote $n=|J|$ as the total number of demands.
To locally service user demand $j \in J_s$, the UAV must reach location $s$ within time window $[r_j,d_j)$ and service at location $s$ for at least a consecutive time $q$.
%wireless communication metric (e.g., SINR, data rate, etc.,) ?
%.%and ensures  sufficient energy to finish servicing the demand.
%The UAV can service an user demand only if it is hovering and if it decides to service a demand, it must hovering at least for a consecutive time $q$ to finish the demand.
If a UAV just reaches the location at time point $t = d_j$, it cannot service demand $j$ but just misses the demand.

%
%At any time, an user demand is referred to as either  \emph{alive}, \emph{serviced} or \emph{missed}.  %to remove
%Initially, the demand is \emph{alive} after being released, indicating that the user is waiting to be serviced, then it becomes \emph{serviced} when a UAV finishes servicing the demand or  \emph{missed} when no UAV services the demand before the waiting deadline of the demand.

In our model, the UAV can simultaneously service multiple user demands at the same location, which holds for many applications such as edge computing and information broadcasting to users \cite{broadcasting2014}.

%Specifically, when the UAV starts to service some user demand at location $s$ at time $t$, any demand $j \in J_s$ satisfying $r_j \le t < d_j$ can be serviced by the UAV simultaneously.
To ensure reliable service,
the servicing demands can not be interrupted and should be finished before the UAV leaves the location.
Note that in our problem of optimizing multi-UAV trajectory planning,
the demand set $J_s$ is already determined as users need to register their demands beforehand and otherwise the demand will not be serviced.
The UAV path planning is computed based on the given $S, J_s, K$ in the offline.
Later in Section~\ref{sec:experiments}, we will show that our designed algorithms also provide good performances even if there is some unexpected error in estimating the demands.

%\footnote{Our analysis and results can be easily extended to the case that it takes a fixed number of time units to complete a service. Similar to \cite{wang2018traffic, xu2018uav, zeng2017energy}. An example of such service is disseminating/broadcasting latest news to users.}.

%Without much loss of generality, we assume the service time for each user demand is short, and as long as we have at least one UAV at location $s$ at time $t$, any demand $j \in J_s$ satisfying $r_j \le t < d_j$ can be serviced by the UAV \footnote{Our analysis and results can be easily extended to the case that it takes a fixed number of time units to complete a service. Similar to \cite{wang2018traffic, xu2018uav, zeng2017energy}, we assume that the UAV can service multiple demands at the same time. An example of such service is disseminating/broadcasting latest news to users.}.
%

%To simplify our analysis, we assume the charging station is a subset of user locations, i.e. $C \subset S$, and restrict that no user demand is released at any charging location.
%.%and charging stations
%
%We consider the users' ground locations in a general graph and
%
We model the spatial connectivity of the UAV locations $S$ by a distance matrix $a(\cdot,\cdot)$, where $a(s,s')$ indicates the pairwise travelling distance from location $s$ to $s'$ with  $s, s' \in S$.
In our model, we assume the UAV moves at constant speed $v$, and without loss of generality, we normalize it as $v = 1$.
In addition, we practically consider that the UAV's traveling distance matrix satisfies the triangle inequality, i.e., for any three different locations $s, s', s'' \in S$, it holds that
\begin{equation}\label{eq:triangle}
a(s,s') + a(s',s'') \ge a(s,s'').
\end{equation}

When a UAV is flying from location $s$ to $s'$, it cannot provide any service to users in the meantime, since UAV-enabled service coverage is small as compared to the distance between location $s$ and $s'$ \cite{8933037}.

%in our model since it is not close to either locations $s$ and  $s'$.

%All input values can be continuous values as our

Our problem can be described as an optimization problem to maximize the number of successfully serviced demands
%or the chance to hit any demand $j$ in its time window $[r_j, d_j)$
as follows.
%| \se{ j~|~[r_j,d_j) \cap T_{s} \not = \emptyset, j \in J_s} |
\begin{equation}\label{eq:obj}
\begin{array}{c}
    \max_{
    \se{T_{k,s}, ~\forall ~ \text{UAV} ~ k ~ \text{and~location}~ s}
    } \limits ~
    \sum_{s \in S} \limits \sum_{
    j \in J_s} \limits \text{serviced}(j, T_s) \\
    s.t. ~(T_{k,1},T_{k,2},\ldots,T_{k,|S|}) \longrightarrow \text{feasible routing}, ~  \forall k. \\
    \se{T_{s}  = \bigcup_{k=1}^K  T_{k,s}, s \in S}
     \longrightarrow \text{collision-free routing}. \\
\end{array}
\end{equation}
%
%&s.t. ~~~~~~T_{s}  = \bigcup_{k=1}^K ~ T_{k,s}, ~~~  \forall s \in S \\
%j \in J_s: [r_j,d_j) \cap T_{s} \not = \emptyset %wrong!
%
%\parbox{.7\textwidth}
\begin{equation}\label{eq:service}
\text{serviced}(j, T_s) =
    \begin{cases}
    1, &
    \text{if}~ \exists ~k, ~ \exists~ [t_1,t_2) \in T_{k,s}, \\
    & t_2 - q \in [r_j,d_j) \cap [t_1,t_2)
    \\
    0,& \text{otherwise}
    \end{cases}
\end{equation}
As a decision, the schedule set $T_{k,s}$  in the whole discrete time horizon $[1, T_{max}]$ contains all the time periods during which UAV $k$ is hovering at location $s \in S$ to service user demands there, and the time periods $(T_{k,1},T_{k,2},...,T_{k,|S|})$ are the result of a feasible routing of a single UAV $k$.
In particular, $T_s$ summarizes the overall servicing time periods at location $s$ by all UAVs.
To address the UAV collision issues, the path planning defined by $T_{k,s}$ is \emph{collision-free} if any two UAVs neither meet at any location nor meet during flying to its target location.
At time point $t \in T_{k,s}$, UAV $k$ can start to service a demand $j \in J_s$ at location $s$ only if this demand is released prior to time $t$ and not missed yet, i.e., $t \in [r_j,d_j)$.
As such, demand $j$ can be serviced if time period $[t_1,t_2) \in T_{k,s}$ of UAV $k$ is long enough to finish the demand, i.e., condition
$t_2 - q \in [r_j,d_j) \cap [t_1,t_2)$ in Eq.~\eqref{eq:service} indicates a service success.
%As such, demand $j$ can be serviced whenever at least one UAV reaches location $s$ during time periods $[r_j,d_j)$, that is, $ [r_j,d_j) \cap T_{s} \not = \emptyset$ in the objective to count as a service success.
Thus, the optimal UAV trajectory planning sends each UAV $k$ to service at each location $s \in S$ appropriately at time periods $T_{k,s}$ such that the total number of serviced demands is maximized.
From above formulation of problem in Eq.~\eqref{eq:obj},
each UAV needs to decide at any time point to visit which location, translating to a huge computational complexity for solving Eq.~\eqref{eq:obj}, which is $O( (T_{max})^{2 K H})$ with $H$ being the number of visits per UAV per location.
Actually, we can rigorously prove that this problem is NP-hard and the existing solutions (e.g., \cite{sugihara2011path, 6198334, xu2020optimized}) in the literature cannot apply, requiring us to innovate and propose new algorithms with provable performance bounds.

\begin{table}[ht!]
\renewcommand{\arraystretch}{1.0}
\caption{Notations and Physical Meanings.}
\label{tbl:notation}
\centering %|p{9cm}|p{6cm}
\begin{tabular}{ l |  p{5.5cm}}
\hline
\hline
{\bf Math notation} & {\bf Physical Meaning}\\
\hline
%$K,S$ & the set of UAVs and Locations respectively \\
$K$ & the set of UAVs \\
\hline
$S$ & the set of user locations \\
\hline
$J, n$ & the set (resp. the number) of all user demands\\
\hline
$J_s$ & the set of user demands at location $s \in S$\\
\hline
$r_j, d_j$ & the release time and waiting deadline of user demand $j \in J$, respectively \\
\hline
$q$ & servicing time for each demand \\
\hline
$a(s,s')$ & the travelling distance from location $s$ to location $s'$\\
\hline
%{\bf Math notation} & {\bf Physical Meaning}\\
%\hline
$U(t)$ & the state of the UAV at time $t$ \\
\hline
$M(t)$ & the set of historical UAV states up to time $t$\\
\hline
$J(M(t))$ & the set of serviced demands up to time $t$ \\
\hline
$\pe{U(t), J(M(t))}$ & the
decision state at time $t$ \\
\hline
\hline
\end{tabular}
\end{table}

\begin{figure}[!t]%[htbp]
\centerline{
\includegraphics[keepaspectratio, width = 0.95 \columnwidth]{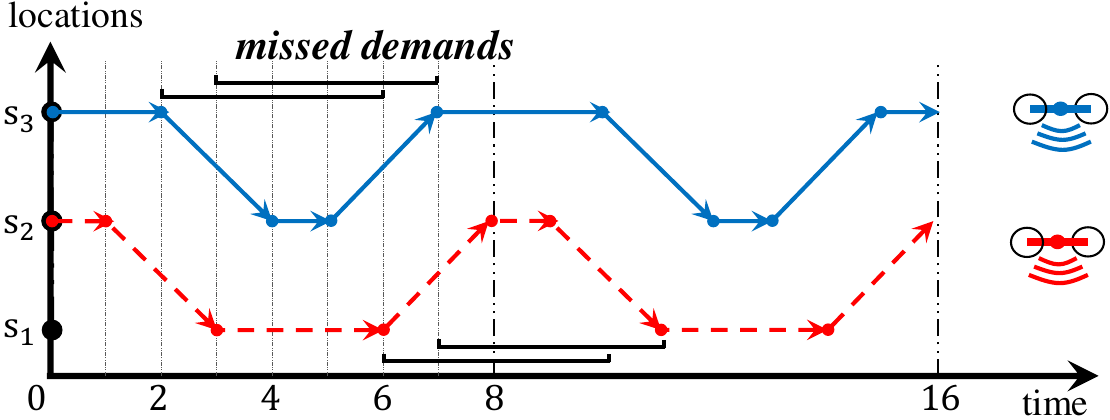}
}
\caption{An illustrative example of two-UAV cooperative path planning (in red dashed line and blue solid line) to service three locations ($s_1, s_2, s_3$) placed on a line over 16 discrete time units.
All demands last for 4 units of time and are released at each discrete time point, i.e.,
$J_{s_1} = \se{[0,4), [1,5), [2,6), ...}$,
$J_{s_2} = \se{[0,4), [0,4), [1,5), [1,5), [2,6),[2,6), ...}$,
$J_{s_3} = \se{[0,4), [1,5),$ $[2,6), ...}$.
The arrows indicate the optimal path planning of the two UAVs, where the horizontal intervals at locations $s_1, s_3$ indicate 4 missed demands.
}
\label{fig-coop}
\end{figure}

%TODO: Difference when including energy:
%more decision for target location

%\KK{routing example needs to change}
%Noise : Modify solution

Seeking UAV-swarm cooperation is important for the optimal path planning for meeting dynamic demand patterns over space and time,
%it is necessary for the UAVs to cooperate in designing their routing jointly,
and we illustrate the cooperation advantage by a simple example in Fig.~\ref{fig-coop}, which motivates our problem formulation in next sections.
Here we have two UAVs to provide service at three locations along a road in set $S = \se{s_1,s_2,s_3}$.
The three locations along a road have 1D coordinates $\se{0,2,4}$, respectively, telling a UAV's travelling delay $a(s_1,s_2) = a(s_2,s_3) = 2$ if we normalize the UAV velocity as one unit per time.
%Once reaching a user location, the UAV can service the user fast with $q \rightarrow 0$ in this example.
At each discrete time $t \in \se{0,1,2,...}$, there is one user demand released at location $s_1$, one user demand released at $s_3$ and two user demands released at middle location $s_2$.
All demands have fixed service time $q = 1$ and last for $4$ time units, i.e., $d_j - r_j = 4, \forall j \in J$.

Fig.~\ref{fig-coop} shows how the two UAVs cooperate with each other in a periodic cycle of $8$ time units at  optimum.
%UAV 1 returns to location $s_1$ (and location $s_2$) in every 8 time units.
Specifically, in every $8$ time units, the two UAVs take turns to visit location $s_2$ with a time gap of $3$ time units
(without missing any demand there), and each of them stays at one of the two ends (location $s_1$ or $s_3$) for $3$ time units for each visit there.
During the first time period cycle $[0,8)$ for example, at location $s_1$ (resp. $s_3$) only two demands with release times $6$ and $7$  (resp. release times $2$ and $3$) are missed.
%, as shown in Fig.~\ref{fig-coop}
That is to say, totally  $87.5\%$ of demands are serviced.
Without cooperation, however, the two
%UAVs stay at their own locations will miss many more demands.
UAVs which stay at their own locations will miss much more demands.
For example, if the first UAV stays at location $s_1$ to meet all the demands there and the second UAV services the remaining two locations, the strategy of the second UAV that
travels between $s_2$ and $s_3$ back-and-forth staying one unit time at each
will miss $6$ demands ($4$ demands from $s_2$ and $2$ demands from $s_3$) in the period time cycle $[0,6)$, which does not service more demands than simply staying at location $s_2$ to service all demands there. As a result, at most $75\%$ of demands can be serviced.
%at most $16$ demands can be serviced by the second UAV,
%resulting in $75\%$ of successfully serviced demands.
Later in Lemma~\ref{lmm:nonshare}, we theoretically prove that the performance of UAV-swarm's partitioned path planning without overlapped locations can be arbitrarily poor.

%When the demand information are completely unknown in advance, no demand can be serviced in any algorithm in the worst case for our problem. One can imagine the case where a new demand always appears at the location where no UAV is nearby and disappears when some UAV arrives the location.
%offline
%In the following, we start with optimizing each UAV given its demands to service. The designed algorithm is the basis of our algorithm for routing multiple UAVs in Section~\ref{sec-multiple-UAVs}.
%lays the foundation of our algorithm for UAV-swarm cooperative routing in Section~\ref{sec-multiple-UAVs}.

%In introduction on page 1, Please cite the following paper:
%\cite{zhang2015data}
%Zhang, Yongmin, Shibo He, and Jiming Chen. "Data gathering optimization by dynamic sensing and routing in rechargeable sensor networks." IEEE/ACM Transactions on Networking 24.3 (2015): 1632-1646.

The advantage of UAV-swarm cooperation is not only to hit the dynamic demands, but also to cover each other to charge battery and enable sustainable service provisioning. Later in Section~\ref{sec:energy} we will generalize our model to consider UAV battery charging at ground stations.

\section{Optimal path planning algorithm for a single UAV} \label{sec:single-drone}
% UAV decisions: start to service a demand, leave a location, charge

In this section, we first design and analyze the optimal path planning algorithm of a single UAV for servicing the dynamically arriving user demands across different spatial locations, which will be shown extendable to the multi-UAV case by taking the cooperation advantage among multiple UAVs into account in later sections. 
Nevertheless, the dynamic optimization problem for the  single UAV is still challenging due to the curse of dimensionality or huge solution space for spatial routing decision-making over time.
%multiple UAVs in Section~\ref{sec:multiple-drone}.
To tackle this challenge, we first analyze the fundamental properties of the UAV routing under optimal path planning, and then identify the critical times at which a routing decision is really necessary. This will substantially reduce the solution space of routing decisions, since now it suffices to make routing actions at a reduced set of time points instead of the whole continuous-time domain. Then, only at each critical time stamp, we need to characterize the \emph{decision state} that determines the optimal routing action.
Finally, we propose a dimensionality-reduced dynamic programming approach to enumerate all such possible decision states and efficiently compute the optimal path planning.

%The optimal path planning contains a series of routing decisions at particular time stamps and it can be regarded as a Markov decision process.
%Therefore, for each routing decision at some time stamp, we needs to characterize the {decision state} that completely determines the optimal routing action.
%We propose an efficient representation for the state of the problem, and based on that we propose a dynamic programming approach to find the optimal path planning for a single UAV.
%
In the following, we first characterize the UAV movement in an optimal path planning by the following two lemmas.
%all the demands that it has started to service
\begin{lemma}\label{lmm:prnc-1}
At the optimum, once the UAV finishes servicing the on-going demands, it should either immediately travel to another location or
continue hovering at the current location until servicing a new demand  here.
\end{lemma}
\begin{proof}
%Lemma~\ref{lmm:prnc-1} can be proved as follows. 
We can prove this by contradiction. If the UAV continues staying at the current location s for a positive time without servicing any demand there, it must be waiting for some new demand to service. Suppose not, then the UAV can always travel to next target location earlier, hence servicing more demands and improving the objective of problem (\ref{eq:obj}).
%it is better to immediately send the UAV to another location for serving positive demands there. 
\end{proof}
%\begin{proof}
%In the opposite case, the UAV will stay at the current location for some period of time before leaving the location, and during this time period  it does not service any demand at all.
%Therefore, it is better for the UAV to leave the location earlier and avoid such meaningless waiting time.
%\end{proof}
\begin{lemma}\label{lmm:prnc-2}
If the UAV just arrives at a new location $s'$ (from location $s$) to just find no demand there, it is optimal for the UAV to keep waiting at location $s'$ until a new demand is released in the future.
%If the UAV arrives at a new user location $s'$ from location $s$ to just find no demand there, it is optimal for the UAV to wait at location $s'$ until a new demand is released to service.
\end{lemma}
\begin{proof}
Consider the other case that the UAV plans to leave location $s'$ to some other location $s''$, without meeting or servicing any demand at location $s'$.
Then, it is better for the UAV to travel directly from location $s$ to $s''$, which saves travel time according to the triangle inequality in Eq.~\eqref{eq:triangle}.
\end{proof}

Using the above two lemmas, we conclude that at optimum the UAV decides to leave (or not leave) the current location only at the time point when it just finishes servicing a demand there.
When the UAV decides to leave the current location at time point $t$, we do not need to make any routing decision during time window $(t-q, t]$, as it takes time $q$ to finish servicing the current demand. 
Therefore, we can equivalently regard this routing decision of leaving or not to be made at time point $t-q$.
\subsection{Dimensionality-reduced decision states in the Markov decision process}\label{subsec:decision-state}
%In Section~\ref{subsec:uav-action}, we characterize the movement of a UAV by three movement states and decompose the routing problem of a UAV into a series of actions (i.e., Fly, Hover, Charge), where each action changes the movement state of a UAV. In addition, we analyze the time events that lead to UAV movement state change.
%
%However, determining the best action at each necessary time stamp requires the current information of the alive demands, location of UAV, remaining energy in battery, etc.
%In this section, we characterize these necessary information as \emph{Decision State} (define later in Definition~\ref{def:decision-state}), which completely determines the optimal action at each time stamp.
%(define later in Definition~\ref{def:decision-state})
The optimal path planning contains a series of routing decisions (e.g., stay at the current location or fly to another location) at critical time stamps, and it can be modeled as a Markov decision process (MDP).
Therefore, we need to characterize the {decision state} that determines the optimal routing action at each critical time point.
A \emph{decision state} consists of the state of the UAV and the state of the demands, both of which are necessary for the UAV to make the best routing decision. For example, before flying to a new location, the UAV should know which user demands can be met and serviced there.
The focus of our dimensionality-reduced dynamic programming approach is to efficiently reduce the possible decision states such that we only consider the decision states that could occur under an optimal path planning.
We successfully achieve this decision state reduction by comparing the path planning choices and disregarding the choices dominated by the others.
Next, we first introduce the UAV state as well as the partial feasible path planning, and then the dominant property over different partial feasible path planning choices. 
Table~\ref{tbl:notation} contains a summary of notations defined below for characterizing the decision state of the problem.

\begin{definition}[UAV State for the MDP formulation] \label{def:uav-state}
at time point $t$, the UAV state is represented by a tuple ${U(t)} = \pe{t,i,s}$, indicating that the UAV is at location $s \in S$ at time point $t$ and it has already serviced a number of $i$ demands up to time $t$ (including the possible demands finished exactly by time $t$).
\end{definition}

%Let $\mathcal{T} = \se{ t~|~ \min_{j \in J} \se{r_j} \le t \le \max_{j \in J} \se{d_j}, t \in \mathbb{N} }$ be the set of times with consideration.
%.//mobihoc2020/mob.tex

\begin{definition}[Partial Feasible Path Planning]\label{def:partial}
%Let $\mathcal{T}$ be a set of time stamps that are necessary for UAV
A \emph{partial feasible path planning} up to time $t$ is a set of UAV states $M(t) = \se{U(t') ~|~ t' \in [0,t] }$,
where $U(t')$ represents the UAV state at time point $t'$.
\end{definition}
%\in M(t)
%\in \se{0,1,...,t}

Before time point $t$, each historical UAV state $U(t') \in M(t)$ contains the location information of the UAV at earlier time $t'$, hence a partial path planning for the UAV up to time $t$ can be defined by set $M(t)$.
As a result, the serviced demands up to time $t$ can also be derived, which is denoted by the set $J(M(t))$ and is critical in our MDP problem objective. 
% when the UAV arrives some location $s$ again, 
%differentiate -> distinguish 
%distinguish -> tell: %
Without recording this set, the UAV cannot tell whether a demand $j$ at location $s$ has been previously serviced or not, by using the basic information of demand release time $r_j$ and deadline $d_j$.
This can indeed happen when demand $j$ has a long waiting time window $[r_j,d_j)$ during which the UAV might have visited location $s$ for several times. 
Each visit after the first one,
%For each such visit except the first, %
the UAV should know that demand $j$ is already serviced at the very first visit. 
%due to the large waiting deadline $d_j$ of demand $j$.
%with only the information of the demand (i.e., release time, deadline).
%Any missed demand from set $J(M(t))$ indicates that this demand can never be serviced by the UAV after time $t$.
However, directly recording this set $J(M(t))$ leads to exponential solution space due to the combinatorial nature of the serviced demands by the UAV over time.
%Note that we only use set $J(M(t))$ to identify 
The key observation in our approach is that the complexity of set $J(M(t))$ can be significantly reduced without the entire information of $M(t)$ but instead with only the latest UAV departure times at the locations.
%, which will be described and used in the next section.
%Next, 
Now, we formally introduce the definition of decision state at a time point $t$  of interest.

%and will be introduced later.
%With the information of UAV $U(t)$ and information of demands  we define the decision state.

\begin{definition}[Decision State] \label{def:decision-state}
In a partial feasible path planning defined by $M(t)$,  the decision state at time point $t$ is represented by $\pe{U(t), J(M(t))}$ where $U(t)$ represents the UAV state at time point $t$ and $J(M(t))$ indicates the set of serviced demands up to time $t$.
\end{definition}

\iffalse
\begin{table}[ht!]
\renewcommand{\arraystretch}{1.0}
\caption{Notations for the states.}
\label{tbl:notation-state}
\centering %|p{9cm}|p{6cm}
\begin{tabular}{ l |  p{10.5cm}}
\hline
\hline
{\bf Math notation} & {\bf Physical Meaning}\\
\hline
$U(t)$ & the state of the UAV at time $t$ \\
\hline
$M(t)$ & the set of historical UAV states up to time $t$\\
\hline
$J(M(t))$ & the set of serviced demands up to time $t$ \\
\hline
$\pe{U(t), J(M(t))}$ & the 
decision state at time $t$ \\
\hline
\hline
\end{tabular}
\end{table}
\fi
%Later in this section we will show how to derive $J(Q(t))$, as well as efficiently encode the decision state $\pe{U(t), J(Q(t))}$.
We then define the dominance relation between two partial feasible path planning choices and show that dominated path planning choices will not be adopted by optimum.

\begin{definition}[Choice Dominance]\label{def:dominanted}
Let sets $M(t)$ and $M^{\#}(t^{\#})$ define any two different partial feasible path planning choices respectively, and let
$U(t) = \pe{t,i,s} \in M(t)$ and
$U^{\#}(t^{\#}) = \pe{t^{\#},i^{\#},s^{\#}} \in M^{\#}(t^{\#})$ be the UAV states at time point $t$ and time $t^{\#}$ in the two path plannings, respectively. Formally, the path planning defined by $M^{\#}(t^{\#})$ is \emph{dominated} by $M(t)$ {\bf if and only if}
(i) $s=s^{\#}, i = i^{\#}$; (ii) $J(M(t)) = J(M^{\#}(t^{\#}))$; (iii) $t < t^{\#}$.
%%\begin{itemize}[label={\textbullet}]
%\begin{enumerate}[label=(\roman*)]
%\item $i = i^{\#}, s=s^{\#}$.
%\item $J(M(t)) = J(M^{\#}(t^{\#}))$.
%\item $t < t^{\#}$.
%\end{enumerate}
%%\end{itemize}
\end{definition}
\begin{lemma}\label{lmm:dominated}
Dominated partial path planning will never be adopted by optimum.
\end{lemma}
\begin{proof}
Comparing the two path plannings defined by $M(t)$ and $M^{\#}(t^{\#})$ in Definition~\ref{def:dominanted}, all state parameters are the same except that $t < t^{\#}$, indicating that when finishing the same objective number $i$ of demands, the former path planning uses less time than the latter and hence has advantage over the latter since the UAV can at least dissipate this extra time via waiting at the current location from time $t$ to time $t^{\#}$ to service more potential demands.
\end{proof}

In the next subsection, we are ready to introduce our dimensionality-reduced dynamic programming algorithm to compute all possible decision states.
%based on their routing actions.

%State Transition
\subsection{Low-complexity dynamic programming algorithm}\label{subsec:dp-transition}

In this subsection, we propose a new algorithm to compute the optimal path planning of a single UAV.
As mentioned in Section~\ref{subsec:decision-state}, the optimal routing decision at each time $t$ can be completely determined by the decision state $\pe{U(t), J(M(t))}$, which corresponds to a partial path planning choice $M(t)$.
We first show a way to represent this decision state efficiently, without $M(t)$.

%without the entire information of $M(t)$.

\begin{lemma}\label{lmm:record-demand}
At time point $t$, it is sufficient to compute the optimal routing action by recording the last departure time $t^{*}$ at each location $s$, instead of recording the set of all serviced demands $J(M(t))$ up to time $t$.
%At the optimum, when the UAV leaves location $s \in S$ at time $t$, any user demand that is released no later than time $t-q$ at location $s$ (i.e., demands from $\se{j \in J_s ~|~ r_j \le t-q}$) is either missed or serviced.
\end{lemma}
%\textit{Sketch of proof.} 
\begin{proof}
Consider the serviced demands at location $s$ up to time $t$, i.e., $J' = J_{s} \cap J(M(t))$. When the UAV just arrives at location $s$ at time point $t$, we focus on an arbitrary demand $j \in J_s$ that is released before time $t$. 
If $d_j \le t$, this demand is already due by time $t$, and it does not matter whether it has been serviced or not. Otherwise if $r_j < t < d_j$, this demand could possibly already be serviced at some earlier visit of location $s$.
% given that time $t^{*}$ is the last departure time at location $s$, 
We claim that demand $j$ is already serviced if $r_j \le t^{*}-q$ and not serviced otherwise.
Note that by lemmas~\ref{lmm:prnc-1} and \ref{lmm:prnc-2}, the UAV leaves the location at a historical time $t^{*}$ only if it exactly finished some demand at time point $t^{*}$. Hence, demand $j$ can at least be serviced by the UAV during historical time interval $[t^{*}-q,t^{*})$ if $r_j \le t^{*}-q$; and otherwise if $r_j > t^{*}-q$, that is, demand $j$ cannot be finished by time $t^{*}$, and thus it is not serviced yet.
%the UAV cannot service demand $j$ at any historical time since this demand cannot be finished by time $t^{*}$ due to the service time $q$.
As a conclusion, with the last departure time $t^{*}$ at location $s$, the UAV is able to differentiate for each demand $j$ from $J_s$ whether the demand is still waiting to be serviced or not.
\end{proof}

\iffalse
\KK{remove below, rewrite above}
%the computation of $\pe{U(t), J(Q(t))}$
Next, we introduce how to derive the set of missed (or serviced) demands $J(M(t))$ from set $M(t)$. We show that only one time stamp is needed to record the serviced or missed demands at each user location.
\begin{lemma}\label{lmm:record-demand-remove}
At the optimum, when the UAV leaves location $s \in S$ at time point $t$, any user demand that is released no later than time $t-q$ at location $s$ (i.e., demands from $\se{j \in J_s ~|~ r_j \le t-q}$) is either missed or serviced.
\end{lemma}
\begin{proof}
According to Lemma~\ref{lmm:prnc-1}, the UAV leaves location $s$ at time point $t$ only if it just services a demand at time point $t$ at location $s$.
In other words, the UAV is servicing at location $s$ during time $[t-q,t)$.
Then, for any demand $j$ that is released no later than $t-q$ at location $s$, it can at least be serviced during time $[t-q,t)$ unless the demand is already missed at time point $t-q$.
Therefore, the lemma is proved.
\end{proof}
\fi

%Lemma~\ref{lmm:record-demand} indicates that the missed or serviced demands at location $s$ can be easily computed by $t-q$ with time $t$ being the latest departure time of the UAV at location $s$.
%Since this departure time corresponds to the completion of some demand, it must be included by some UAV state $U(t') \in Q(t)$ and hence can be computed efficiently.
Lemma~\ref{lmm:record-demand} indicates that whether a demand from location $s$ is waiting to be serviced or not can be easily identified by the last UAV departure time $t^*$ at location $s$.
This provides opportunities for efficiently encoding the decision state $\pe{U(t), J(M(t))}$, without the entire information from set $M(t)$.
Based on Lemma~\ref{lmm:record-demand}, we now describe the dimensionality-reduced dynamic programming algorithm as follows.

At time point $t$, we label the latest UAV departure time stamps at all the locations.
%, as they are sufficient to identify the missed or serviced demands up to time $t$,  as shown in Lemma~\ref{lmm:record-demand}.
%to label the UAV departure times
More specifically, we introduce set $Q$, where each tuple $\pe{s^{*},t^{*}} \in Q$ tells that time point $t^{*}$ is the latest time when the UAV leaves location $s^{*}$. 
%Let $J(Q)$ be the corresponding missed or serviced demands computed from  set $Q$ (according to Lemma~\ref{lmm:record-demand}).
%
Therefore, the decision state at time point $t$ can be efficiently represented by $\pe{Q,i,s,t}$, which indicates that the UAV is at location $s$ and  just services an accumulated number $i$ of demands by time $t$ with the constraints of departure times in $Q$.
%missed or serviced demands $J(Q)$.
%
Although set $Q$ still incurs huge solution space, we only need it to extract useful information for optimal decision-making.
Later in the analysis of computational complexity of our dimensionality-reduced dynamic programming algorithm, we show that the complexity of set $Q$ can be significantly reduced to achieve polynomial complexity.
%and later we show that $J(Q)$ can be encoded efficiently.

%focus on the actions that transits the state to new state.
Based on the UAV decision state $\pe{Q,i,s,t}$, we further consider the dominance relation between different states.
Implied by the dominant property of decision state in Definition~\ref{def:dominanted} and Lemma~\ref{lmm:dominated}, the earlier the UAV services the same demands, the better the path planning choice is.
We thus aim to design the dimensionality-reduced dynamic programming algorithm to find the path planning choice that can service the same amount of demands as the optimum and minimize the time span to service these demands.
Accordingly, given particular set $Q$, integer $i \in \se{1,...,n}$ and location $s \in S$, we denote variable $g(Q,i,s)$ as the earliest time $t$ such that some partial feasible path planning reaches decision state $\pe{Q, i,s,t}$, i.e., time $t$ is the earliest time for the UAV to finish $i$ demands under the constraints of departure times $Q$.
%We restrict our consideration such that
By this definition, a user demand is indeed serviced exactly at time point $t$, which holds for the optimum.
The maximum value $i$ such that some state $\pe{Q,i,s,t}$ corresponds to a partial feasible path planning choice indicates an optimal solution to our problem.
%(i.e., $g(Q,i,s) \not = \infty$)
The dimensionality-reduced dynamic programming formulation above has the advantage that when multiple optimal solutions exist, it always returns the one with the shortest time to service the same objective number of demands.
This is consistent with practice that the UAV can stay in the air for a limited amount of time.

%tiny,scriptsize,footnotesize,small,normalsize

\begin{algorithm}[!hbt] \small
\caption{Optimal path planning of a single UAV. }
\label{alg:one-general}
%\begin{multicols}{2}
\begin{algorithmic}[1]

\Require set $J$ of demands, set $S$ of locations.

\Ensure optimal path planning.

%$\textit{Computed} \gets \emptyset$
\State $g(\cdot) \gets \MX$, $G(\cdot) \gets \emptyset$
%\Comment{{ Priority Queue $G$}}

%Insert $\pe{\emptyset, \mu(s,\min(R_s)), s, \min(R_s)}$ into $G$

\For{$s \in S$}
\Comment{{ Initialization}}

\State
$t \gets \min \se{r_j ~|~j \in J_s}$
\State
$i \gets \sum_{j \in J_{s}: r_j = t} 1$
\State
$g(\emptyset, i, s) \gets t + q$
\State
{Insert} $\pe{\emptyset, s}$ {into} $G(i)$
\EndFor

%\For{$i \in \se{1,2,...,n}$}
\For{$i = 1$ to $n$}
\Comment{{ Round $i$}}
\label{alg-line-roundi}

\For {$\pe{Q, s} \in G(i)$}
\label{alg-line-loop2}
%{$G \not = \emptyset$}

%\State Pull $\pe{Q, i, s, t}$ from $G$ with smallest $t$.
%\label{alg-line-extract-min}

%\State ${g}(Q,i,{s}) \gets t$
\State $t \gets {g}(Q,i,{s})$
%\State $\textit{Computed} \gets \textit{Computed} \cup \se{(Q,i,{s})}$.

\For{$s' \in S$}
\label{alg-line-loop3}
\State
$\pe{Q',i',s',t'} \gets \textsc{Expand}(\pe{Q,i,s,t}, s')$
\label{alg-line-expand}

\State
$Q' \gets \textsc{Trim}(\pe{Q', i',s',t'})$
\label{alg-line-trim}

%\If{$(Q',i',s') \not \in \textit{Computed}$}

\State $g(Q',i',s') \gets  \min \se{g(Q',i',s'), t'}$
\label{alg-line-update}

\State
Insert $\pe{Q',s'}$ into $G(i')$
%\EndIf

\EndFor

\EndFor
\EndFor

\State
$Q^{*},i^{*},s^{*} \gets \argmax_{i} \se{g(Q,i,s) \not = \infty}$
\label{alg-line-opt}

\State
path planning $\gets$ backtrack from $g(Q^{*},i^{*},s^{*})$
%%%%%%%%%%%

\;
%\\\hrulefill
\Procedure{Expand}{$\pe{Q, i,s,t}$, $s'$}
\Comment{{ Expand}}

\If{$s = s'$}
\label{alg-line-stay}
\State
$t^{*} \gets \min \se{r_j ~|~ t - q < r_j, j \in J_{s'}}$
\State
$Q' \gets {Q}$
\State
$ i' \gets i + \sum_{j \in J_{s'}: r_j = t^{*}} 1$
%\ElsIf{$\exists \pe{s',t^{*}} \in Q$}
\State
{\bf Return} $\pe{Q', i',s',t^{*} + q}$

\Else

\State
$t' \gets t + a(s,s')$
\label{alg-line-upt}
\State
$Q' \gets Q \cup \se{\pe{{s},t}}$
\label{alg-line-upq}

\If {$\exists~\pe{s',t^{*}} \in Q$}
\State $i' \gets i + \sum_{j \in J_{s'}: t^{*} - q < r_j \le t' < d_j} 1$
\label{alg-line-move-i}
%$Q' \gets (Q \setminus \se{\pe{s',t^{*}}}) \cup \se{\pe{{s},t}}$
\Else
\State $i' \gets i + \sum_{j \in J_{s'}: r_j \le t' < d_j} 1$
\label{alg-line-move-ii}
\EndIf

%\If{$i'=i$}
\State
	{\bf Return} $\textsc{Expand}(\pe{Q',i',s',t' + q}, s')$
	{\bf If} $(i'=i)$.
	\label{alg-line-move-iii}
%\EndIf
\State
{\bf Return} $\pe{Q', i',s',t' + q}$
\label{alg-line-move-ret}
\EndIf

\EndProcedure

%%%%%%%%%%%

\;
%\\\hrulefill
\Procedure{Trim}{$\pe{Q,i,s,t}$}
\label{alg-line-trimcond}
\Comment{{ Trim}}

\State $Q^{*} \gets \emptyset$
\For{$\pe{s^*,t^*} \in Q$}

%\State
%\parbox{0.85\linewidth}
%{ {\bf Continue If} $s^* = s$  {\bf or} $\exists \pe{s'',t''} \in Q, s^{*} = s'', t^{*} < t'' $  {\bf or} $t^{*} - q + \max_{j \in J_{s^*}} (d_j-r_j) \le t + a(s,s^{*}) $}

%\newline
\If{{$s^* = s$  {\bf or} $\exists \pe{s'',t''} \in Q, s^{*} = s'', t^{*} < t'' $ {\bf or} $t^{*} - q + \max_{j \in J_{s^*}} (d_j-r_j) \le t + a(s,s^{*}) $}}
\State {\bf continue}
\EndIf
%$Q \gets Q \setminus \se{\pe{s^*,t^*}}$
\State  $t' \gets \max \se{r_j ~|~ r_j \le t^* - q, j \in J_{s^{*}}}$
\State $Q^* \gets Q^* \cup \se{\pe{s^*,t' + q}}$
\EndFor

\State {\bf Return} $Q^*$
\EndProcedure

\end{algorithmic}
%\end{multicols}
\end{algorithm}

%Algorithm~\ref{alg:one-general} shows the computation of variables $g(Q,i,s)$, which adopts the main concept of dijkstra's algorithm for shortest path.
Algorithm~\ref{alg:one-general} shows how to compute all variables $\se{g(Q,i,s)}$ for the dimensionality-reduced dynamic programming.
%
%Time $g(Q,i,s)$ can be expressed in an inductive form.
Suppose $g(Q,i,s) = t$ is computed, by definition some user demand is serviced exactly at time point $t$.
This is interpreted as a decision process that the UAV decides at time point $t-q$ to start servicing that demand, i.e., $t-q$ is regarded as a \textit{decision-making time stamp}.
We enumerate all possible subsequent decisions after time $t-q$ to obtain new partial path planning choices.
Specifically, we aim to obtain a new partial path planning $\pe{Q',i',s',t'}$ from current $\pe{Q,i,s,t}$, by determining the next location $s'$ for the UAV to visit after time $t - q$.
We enumerate all such possible choices and divide them into following two cases.
%Similar to Algorithm~\ref{alg:short}, the algorithm runs in many rounds where in each round $i \in \se{1,2,...,n}$, we consider the feasible solutions that finishes a number $i$ of demands.
%For each computed $g(Q,i,s)$, we extract a feasible solution $\pe{Q,i,s,t}$, providing that $g(Q,i,s) = t$, i.e., the minimum possible time to finish $i$ demands with parameters $Q$ and $s$.
%We aim to extend the partial solution to a new feasible solution $\pe{Q', i',s',t'}$ by guessing the next location $s'$ to be visited after time $t$.

\begin{itemize}
\item
After time $t - q$, the UAV plans to stay at the same location $s$ to service more demands, i.e., $s' = s$ (see Line~\ref{alg-line-stay} of Algorithm~\ref{alg:one-general}). By Lemma~\ref{lmm:prnc-1} it stays at least until time $t' = t^{*} + q$ with $t^{*} = \min \se{r_j ~|~ t - q < r_j, j \in J_s}$ being the earliest time that the next demand(s) is released at location $s$ after time $t - q$.
Parameter $Q'$ is updated to be the same as $Q$ about departure time stamps, since the UAV does not leave the current location.
And the formula $\sum_{j \in J_{s'}: r_j = t^{*}} 1$  contains the additional demand to be serviced at time point $t^{*}$ at location $s'$.
\item
Otherwise after time $t - q$, the UAV decides to leave location $s$. Then it leaves at time point $t$ when finishing servicing the demand. 
%That is to say, the decision of leaving location $s$ is made at time $t-q$.
During time interval $(t-q,t]$, the UAV cannot start to service any new demand at location $s$, otherwise they will not be finished by time $t$.
After time $t$, the UAV will arrive the new location $s'$ at time point $t' = t + a(s,s')$ (see Line~\ref{alg-line-upt} of Algorithm~\ref{alg:one-general}).
To characterize the decision state at time point $t'$, we further divide this case into two sub-cases, depending on whether location $s'$ has been previously visited or not.
\begin{itemize}
\item
If $s'$ is visited before by the UAV, there exists $\pe{s',t^{*}} \in Q$ indicating that the UAV has lastly visited location $s'$ at a previous time $t^{*}$. Note that it takes time $q$ to service a demand, and those demands arriving in time interview $(t^* - q, t^*]$ cannot be serviced at location $s'$. 
Then the number of new demands that can be serviced by future time $t'$ is $\sum_{j \in J_{s'}: t^{*} - q < r_j \le t' < d_j} 1$ (Line~\ref{alg-line-move-i} of Algorithm~\ref{alg:one-general}), 
where only a demand $j$ with release time $r_j$ later than $t^*-q$ and before $t'$ will be considered by time $t'$, and demand $j$ should have deadline $d_j$ longer than $t'$ (i.e., not missed). 
%which excludes the demands that are already serviced before time $t^{*}$.
%that are  serviced by the UAV
%previously at time $t^{*}$ or earlier.
%This case is the major difference with Algorithm~\ref{alg:short}.
\item
If not, $\sum_{j \in J_{s'}: r_j \le t' < d_j} 1$ (Line~\ref{alg-line-move-ii} of Algorithm~\ref{alg:one-general}) tells the number of demands that can be serviced by future time $t'$.
% at location $s'$.
%, which includes all demands at location $s'$ that can be serviced at time $t'$.
\end{itemize}

%\indent
For both sub-cases, the UAV stays at location $s'$ at least until time $t'+q$ to finish the demands started servicing at time point $t'$ (Line~\ref{alg-line-move-ret}). We also update set $Q$ by including one more tuple $\pe{s,t}$, telling that the UAV departs location $s$ at time point $t$ (Line~\ref{alg-line-upq}).
Moreover, if the UAV finds no new demand at future time $t'$, i.e., $i' = i$ (Line~\ref{alg-line-move-iii}), by Lemma \ref{lmm:prnc-2} the UAV should stay at location $s'$ until a new demand is released.
Then, we further expand the partial path planning to find the earliest demand to service at location $s'$ after time $t'$.
%%%%%%In this situation, we first restrict that the UAV stays for at least an amount of $q$ time and recursively expand the partial path planning to find the earliest demand to service.
\end{itemize}

\noindent
Using the above procedure of extending partial path planning solutions to future critical time stamps, we are able to compute all variables $\se{g(Q,i,s)}$ in the whole time horizon.
%
%since all time labels in the algorithm are only used to count the number of serviced demands.
% and to identify the next released demand
%Later in this section, we analyze the computational complexity of Algorithm~\ref{alg:one-general}.
Next, we first prove the optimality of Algorithm~\ref{alg:one-general} and then analyze its computational complexity.  
%Due to the page limit, the complete proof of Proposition~\ref{prop:opt-one-general} can be found in our online technical report \cite{wang:2020}.

\begin{proposition}\label{prop:opt-one-general}
Algorithm~\ref{alg:one-general} computes an optimal path planning choice for the UAV in the whole time horizon.

%Algorithm~\ref{alg:one-general} computes all feasible variables $\se{g(Q,i,s)}$ correctly and hence returns an optimal path planning choice for the UAV in the whole time horizon.
\end{proposition}
\begin{proof}
In Algorithm~\ref{alg:one-general},  decisions are only made at the time $t-q$ when the UAV starts to service a demand at some location $s$. This decision indicates that either after finishing the demand the UAV will leave the location at time $t$ or the UAV continues servicing more demands at the current location after time $t-q$.
For the former case, we enumerate all possible locations that the UAV is going to visit after time $t$ and apply Lemma~\ref{lmm:prnc-2} to make sure that at least one demand will be serviced at the new location after the arrival of the UAV.
For the latter case, we apply Lemma~\ref{lmm:prnc-1} to make sure that at least one more new demand will be serviced by the UAV at the current location.
For both possible decisions, at least one demand will be serviced in the near future, hence the next decision-making time can also be calculated.
%Any other decisions are not considered in the algorithm since they will not be adopted by the optimal path planning according to Lemma~\ref{lmm:prnc-1} and Lemma~\ref{lmm:prnc-2}.

Regarding the optimality achieved by Algorithm~\ref{alg:one-general}, on one hand, for any partial solution $\pe{Q',i',s',t'}$ expanded from $\pe{Q,i,s,t}$, we follow the rule that $t' > t$ and $i' > i$.
That is to say, given that the UAV finishes some demand at time $t$ at location $s$, time $t'$ is the time when the next demand will be finished (at location $s'$) by the UAV.
%On the other hand, we have tested every possibility of next location $s'$ that the UAV might visit after time $t$.
On the other hand, our dynamic programming algorithm enumerate all possible subsequent decisions for the UAV as described above, at least one decision will be exactly the same as the optimum.
Therefore, by induction, $g(Q',i',s')$ is computed correctly assuming that any $g(Q,i,s)$ with $i < i'$ is computed correctly.
This can be seen in Line~\ref{alg-line-roundi} of Algorithm~\ref{alg:one-general} that value $i$ is enumerated increasingly from $1$ to $n$.

As a consequence, the maximum vale $i^*$ such that $g(Q^*,i^*,s^*) \not = \infty$ (Line~\ref{alg-line-opt} of Algorithm~\ref{alg:one-general}) for some $Q^*$ and some location $s^* \in S$ indicates the maximum number of demands that the UAV can service in the optimal path planning.
\end{proof}

%Algorithm~\ref{alg:one-general}

%This can be seen in Algorithm~\ref{alg:one-general} with two folds:
%i) we have tested every possibility of the next location $s'$ to be visited by the UAV after time $t$ when expanding the partial solution $\pe{Q,i,s,t}$.
%ii) the partial solution $\pe{Q,i,s,t}$ extracted from the priority queue (Line~\ref{alg-line-extract-min}) has the smallest value $t$.

Finally, we analyze the computational complexity of Algorithm~\ref{alg:one-general}.
Generally, the computational complexity relates to the number of variables $g(Q,i,s)$ and the complexity for computing them.
%In the following, we first introduce our approach to further reduce the space complexity for encoding the set $Q$, which corresponds to Line~\ref{alg-line-trim} of Algorithm~\ref{alg:one-general}.
%Although the awaiting demands (i.e., not serviced yet) at one location can be identified by a time stamp according to Lemma~\ref{lmm:record-demand}, recording them at all locations still exponentially depends on the number of locations.
%Therefore, to make our algorithm tractable, we define a parameter $\alpha$ such that the algorithm only needs to record for at most $\alpha$ locations.
The complexity of our Algorithm~\ref{alg:one-general} still depends on the mobility of the UAV during a demand window as defined below. 
\begin{definition}\label{asump-demand}
%Let $\alpha \in [1, |S|]$ be a constant. The maximum number of different locations that the UAV can visit during the maximum waiting time $\max_{j\in J} d_j-r_j$ among all the demands is $\alpha$.
Let $\alpha \in \se{1,..., |S|}$ be the maximum number of different locations that a UAV can visit during the longest waiting time $\max_{j\in J} d_j-r_j$ among all the demands in set $J$.
\end{definition}
%%Kai: make it locally relevant

%$Besides $\alpha$,
Given the fixed mobility parameter $\alpha$, the computational complexity of Algorithm~\ref{alg:one-general} only depends on the number of demands and the number of locations, instead of the tedious input values of times (e.g., release times) and distances in traditional dynamic programs.
This is achieved by further trimming set $Q$ (see Line~\ref{alg-line-trim} of Algorithm~\ref{alg:one-general}) to reduce its complexity, which is given in the following Proposition~\ref{prop:time-one-drone}.
%Due to the page limit, the complete proof of Proposition~\ref{prop:time-one-drone} can be found in our online technical report \cite{wang:2020}.

\begin{proposition}\label{prop:time-one-drone}
Algorithm~\ref{alg:one-general} has low computational complexity $O(n^{\alpha} |S|^{\alpha+1})$, which is polynomial in both demand number $n$ and location number $|S|$.
\end{proposition}

\begin{proof}
As Algorithm~\ref{alg:one-general} has three loops (Line~\ref{alg-line-roundi}, \ref{alg-line-loop2}, \ref{alg-line-loop3}), the overall computational complexity is $O(Q) \cdot O(n |S|^2)$, where $O(Q)$ is the space complexity for set $Q$ and $O(n |S|^2)$ is the computational complexity for computing each variable $g(Q,i,S)$. In the following, we show that $O(Q) = O(n ^{\alpha-1} |S| ^{\alpha-1})$.

Although there are arbitrarily many choices for set $Q$, we only focus on those which are necessary for computing the optimal path planning.
This is achieved by applying a trimming process on set $Q$ (Line~\ref{alg-line-trim}).
Specifically, given a partial path planning $\pe{Q,i,s,t}$, we apply a trimming process on set $Q$ as follows (Line~\ref{alg-line-trimcond}).

\begin{enumerate}[label = (\roman*)]
\item%(i)
We remove any tuple $\pe{s^{*},t^{*}}$ from $Q$ if $s^{*} = s$ {\bf or} there exists $\pe{s'',t''} \in Q$ such that $s^{*} = s'', t^{*} < t''$ {\bf or} $t^{*} - q + \max_{j \in J_{s^{*}}} (d_j-r_j) \le t + a(s,s^{*})$.
\item %(ii)
We replace each $\pe{s^{*},t^{*}} \in Q$ by $\pe{s^{*},t'}$, where $t' - q$ is the largest demand release time such that $t' \le t^{*}$, i.e. $t' - q = \max \se{r_j ~|~ r_j \le t^{*} - q, j \in J_{s^{*}}}$.
\end{enumerate}

Note that set $Q$ is only used to count the number of serviced demands, which happens only at Line \ref{alg-line-move-i}.
The above trimming process is designed to make sure that they will not affect the result in Line \ref{alg-line-move-i}.
As we mentioned earlier, if the UAV repeatedly visits a location $s^{*}$, we only label the latest departure time $t^{*}$.
This can be achieved by trimming condition $s^{*} = s$ and condition $s^{*} = s'', t^{*} < t''$ in step (i).
For condition $s^{*} = s$, there is no need to record the latest departure time at location $s^{*}$ since currently the UAV is already at location $s^{*}$.
For condition $s^{*} = s'', t^{*} < t''$, the latest departure time at location $s^{*}$ is $t''$ instead of $t^{*}$.
%
%Assume the UAV has visited some location $s^{*}$ at an earlier time $t^{*}$ (i.e., $\pe{s^{*},t^{*}} \in Q$) and now it is at location $s$ at time $t$, aiming to finish the future demands after time $t$.
%for feasible solution $\pe{Q, i,s,t}$,
%If the UAV visits the same location $s^{*}$ twice, we only take the one with the latest visit, this corresponds to condition $s^{*} = s$ and condition $s^{*} = s', t^{*} < t'$ in step (i).

Moreover, when the UAV arrives location $s^{*}$ in the near future, it has to differentiate whether the demand there has been previously serviced or not since we do not explicitly record the serviced demands in the algorithm.
If $t^{*} - q + \max_{j \in J_{s^*}} (d_j-r_j) \le t + a(s,s^{*})$, we can find that any (already serviced) demand $j$ released before time $t^{*} - q$ at location $s^*$ will be due before the UAV visits location $s^{*}$ at the earliest possible time $t + a(s,s^{*})$.
%, i.e., demand $j$ will not be hit by the UAV twice.
That is to say, there is no need for the UAV to differentiate the serviced demands there since they will be due before the UAV arrival.
%when the UAV arrives location $s^{*}$ in the near future, any previously serviced demand will be already due.
%any demand that can be serviced is not considered by the UAV before.
Therefore, we do not need to record the last departure time $t^{*}$ at
location $s^{*}$, i.e.,  $\pe{s^{*},t^{*}}$ can be removed from $Q$.

On the other hand, in step (ii), if $t^{*} - q$ is not a demand release time, no demand is released during $(t'-q,t^{*}-q]$ at location $s^{*}$, where $t' - q = \max \se{r_j ~|~ r_j \le t^{*} - q, j \in J_{s^{*}}}$. Hence, without affecting counting the number of serviced demands at  Line~\ref{alg-line-move-i} in Algorithm~\ref{alg:one-general}, we could replace $\pe{s^{*},t^{*}}$ by $\pe{s^{*},t'}$ and note that $t'-q$ is a demand release time.
%the number of possible value for time $t'$ here is bounded by $O(n)$.
After step (i), there are at most $\alpha-1$ elements in set $Q$ by definition of $\alpha$, and after step (ii) the time labels in set $Q$ are demand release times where each can be represented in $O(n)$ space.
As a result, after applying the trimming process, the space complexity for set $Q$ is $O( n^{\alpha-1} |S|^{\alpha-1})$, which yields overall computational complexity $O(n^{\alpha} |S|^{\alpha+1})$.
This completes the proof.
\end{proof}
%\fi

%In term of computational complexity, our  Algorithm~\ref{alg:one-general} do not depend on the length of the time horizon but the number of demands and locations.

%\begin{proposition}
%Algorithm~\ref{alg:one-general} returns the optimal routing for each UAV and it has computational complexity $O(n^{\alpha} |S|^{\alpha+1})$, which is polynomial to the demand number $n$ and location number $|S|$.
%\end{proposition}
%\begin{proof}
%We show that there are at most $O(n^{\alpha} |S|^{\alpha})$ necessary feasible solutions.
%\end{proof}
%As Section~\ref{sec-short-demand} is a special case of $\alpha=1$, the complexity $O(n^{\alpha} |S|^{\alpha+1})$ reduces $O(n |S|^2)$ at $\alpha = 1$.

\section{Different user service times}\label{sec:diff_user}
In this section, we generalize our proposed Algorithm~\ref{alg:one-general} to consider flexible user service time. Specifically, for each user demand $j \in J$, its service time is now $q_j$, instead of fixed time $q$.
We show that Algorithm~\ref{alg:one-general} can be generalized to compute the optimal path planning with the following assumption on UAV service.
\begin{assumption}[Irrevocable UAV decision]
When the UAV decides to leave location $s$ at time point $t^{*} = t - q_{j'}$ for some demand $j' \in J_s$ (i.e., leaves at time point $t$), for any demand $j \in J_s$ that is released before time $t^{*}$ but neither serviced nor missed yet, either 
i.) the UAV services this demand before leaving, or
ii.) demand $j$ is no longer  serviced by the UAV at any future visit of location $s$, i.e., demand $j$ is missed.
\end{assumption}

The above assumption deals with the situation that the UAV rejects to service an already-released demand $j$.
Previously, in the case of fixed service time $q_j = q$, condition i.) will always hold because anyway demand $j$ can be finished by the time $t$ when the UAV leaves.
However, for the case of flexible service time, demand $j$ may have very long service time. In that situation, the UAV could possibly service it in future visit, while the above assumption prohibits this to happen. 
This assumption is consistent with many dynamic planning situations where decisions are irrevocable, i.e., when the UAV rejects to service a demand for the first time, it cannot service it in future \cite{lee2018online}.

With the above assumption, we present the generalization of Algorithm~\ref{alg:one-general}.
Firstly, the important resultant observation by the above assumption is that, any demand that is released before time point $t^{*}$ at location $s$ is either serviced or missed by time point $t$ when the UAV leaves, which is consistent with the previous case in Algorithm~\ref{alg:one-general}.
Previously, Algorithm~\ref{alg:one-general} records the latest departure time $t$ at each location $s$, indicating that all demands that are released before time $t-q$ at location $s$ is handled (i.e., either serviced or missed), and hence the UAV only checks demands released after time $t-q$ for future visit at location $s$ (see Line~\ref{alg-line-move-i} of Algorithm~\ref{alg:one-general}). %
In the generalized algorithm, we alter Algorithm~\ref{alg:one-general} to record $\langle t, q_{j'}\rangle$ instead of $t$ in the problem state, i.e., 
the problem state is updated as $\pe{Q,i,s,\pe{t, q_{j'}}}$, instead of $\pe{Q,i,s,t}$.
%, and correspondingly we use new variable $g(Q,i,q_{j'}, s)$.
%, where the latest departure time point $t$ and the corresponding service time $q_{j'}$ for the latest serviced demand $j' \in J_s$.

%As a result, the complexity of encoding set $Q$ will increase by a factor of $O(n^2)$.
%In particular, for each $\pe{s,\langle t, q_{j'}\rangle} \in Q$, on one hand including $q_{j'}$ takes additional $O(n)$ space, on the other hand, time point $t$ now has complexity $O(n^2)$ due to $t-q_{j'} = r_j$ for some demand release time $r_j$, instead of $t - q = r_j$ previously.
%%the time point $t$ maps to a demand release time, i.e., $t - q = r_j$, but now it is $t-q_{j'} = r_j$, implying that the space complexity of time point $t$ is $O(n^2)$, instead of $O(n)$.
%%
%%Moreover, the problem state is updated as $\pe{Q,i,s,\pe{t, q_{j'}}}$, instead of $\pe{Q,i,s,t}$, which incurs another additional $O(n)$ complexity.
%%For each computed $g(Q,i,s)  = t$, by definition, some user demand is serviced exactly at time point $t$, hence, in the new algorithm, to record such demand $j'$ we use new variable $g(Q,i,q_{j'}, s)$.
%In summary, the complexity of the generalized algorithm will increase by a factor of $O(n^3)$ due to including $q_{j'}$.

We update $q$ properly in  Algorithm~\ref{alg:one-general}, and show that the resulting algorithm still returns the optimal path planning.
At the current decision time point $t-q_{j'}$ with problem state $\pe{Q,i,s,\pe{t, q_{j'}}}$, the UAV is planned to service demand $j'$ during time period $[t-q_{j'},t]$ at location $s$, 
we compute for two possible future UAV decisions after time point $t$, either staying at location $s$ or flying to other location $s'$.
In particular, all future released demands that can be finished by time point $t$, i.e., $\se{j~|~ t - q_{j'} \le  r_j \le t - q_j, j \in J_s}$, will all be serviced by the UAV, and hence will be exempt from consideration. 

If the UAV decides to stay at the current location, 
it waits until a future demand $j \in J_s$ release such that $r_j + q_j > t$.
Note that demand $j$ is not necessary the earliest released demand after time point $t - q_{j'}$, since the UAV now may skip some demand of long service time to achieve optimality.
Therefore, similarly as previous we enumerate such future demand $j$ in the new algorithm and update $t \gets r_j + q_j$ and $q_{j'} \gets q_j$ for new problem state.
Otherwise, the UAV leaves at time point $t$ and travels to some other location $s'$. Similarly, we add the departure information  $\pe{s,\pe{t,q_{j'}}}$ at location $s$ to set $Q$.
When the UAV arrives at location $s'$ at time point $t_1 = t + a(s,s')$, we compute the next demand $j$ can be serviced at location $s'$.
Especially, among the demands, say $J_1$, that are available at time point $t_1$, the algorithm enumerates demand $j$ from $J_1$ and transits to next decision time point by  $t \gets t_1+q_j, q_{j'} \gets q_j$.
As such, the UAV will stay at location $s'$ during time period $[t_1,t_1+q_j]$ to service demand $j$, as well as any demand from $J_1$ that can be serviced by time $t_1+q_j$.
To compute $J_1$, if $\exists \pe{s',\pe{t',q_{j''}}} \in Q$, i.e., the latest historical UAV departure information at location $s'$ is recorded in set $Q$, $J_1 = \se{j~|~t'-q_{j''} \le r_j \le t_1 < d_j, j \in J_{s'}}$ and the algorithm only deals with demands released after time point $t'-q_{j''}$.
Otherwise, $J_1 = \se{j~|~ r_j \le t_1 < d_j, j \in J_{s'}}$.
In case set $J_1$ is empty, similarly, the UAV waits until a new demand released at location $s'$.
%
%
%(more details)
%In Algorithm~\ref{alg:one-general}, the current decision time point $t-q$ at Line~\ref{alg-line-stay} corresponds $q = q_{j''}$; the historical decision time point  $t^{*}-q$ at Line~\ref{alg-line-move-i} corresponds $q = q_{j'}$ for $\exists \pe{s', \langle t, q_{j'}\rangle } \in Q$; the future decision time point $t+q$ at Lines \ref{alg-line-move-iii} and \ref{alg-line-move-ret}  corresponds $q = q_{j'''}$ where the UAV starts to service demand $j'''$ at time point $t^{*}$ and demand $j'''$ has the longest service time.
%By such generalization, expanding $\pe{Q,i,s,\pe{t,q_{j''}}}$ to $\pe{Q',i',s',\pe{t',q_{j'''}}}$ still maintains the properties that both time points $t-q_{j''}$ and $t' - q_{j'''}$ are decision-making time points for the UAV, which is consistent with Algorithm~\ref{alg:one-general}.
%Finally, the complexity of Algorithm~\ref{alg:one-general} will only increase by a factor of $O(n^2)$ due to including $q_j$ in $Q$.

The complexity of encoding the new set $Q$ is $O(n^{2(\alpha-1)} |S|^{\alpha-1})$.
For each departure  $\pe{s,\pe{t,q_{j'}}}$ recorded in $Q$, it takes $O(n^2|S|)$ complexity where 
time point $t$ still has complexity $O(n)$ due to $t-q_{j'} = r_j$ for some demand release time $r_j$ ( where it is $t - q = r_j$ previously).
Each computation of variable $g(\cdot)$ increases by a factor of $O(n)$ for computing the next demand $j$ to service since now the algorithm may skip some demand of long service time.
%$O(n^2|S|^2)$
As a result, the new algorithm has complexity $O(n^{2\alpha} |S|^{\alpha+1})$.

\begin{proposition}\label{prop:time-one-drone-flexible}
For different user service time case,
Algorithm~\ref{alg:one-general} can be generalized to compute the optimal path planning within computational complexity $O(n^{2\alpha} |S|^{\alpha+1})$.
\end{proposition}

\section{
Cooperative path planning algorithms for the UAV-swarm } \label{sec:multiple-drone}

In this section, we study the more complex problem of a number $|K|$ of UAVs' cooperative path planning for jointly serving $|S|$-many locations with nontrivial $|K| \ge 1, |S| \ge 2$.
Without loss of generality, we assume the number the UAVs is less than the number of locations, i.e., $|K| < |S|$, otherwise the problem becomes trivial since 
the optimal solution simply assigns at least one UAV to each location and not miss any demand there. Due to the problem difficulty, we focus on approximation algorithm design.
In the following, we first show that the dimensionality-reduced dynamic programming approach in Algorithm~\ref{alg:one-general} is no longer tractable with a large number of UAVs here. 
Afterwards, we provide a common-sense benchmark approach by partitioning UAV-swarm to route and service without any overlapping in locations, as explained in Fig.~\ref{fig:partition} later. As the benchmark performance will be proved to be arbitrarily poor in the worst case, we finally propose a desirable algorithm with constant approximation ratio 
to solve the difficult problem efficiently. %Moreover, we show that any partition-based solution has approximation ratio no more than $\frac{1}{|K|} + \frac{|K|-1}{|S|}$, by a constructive example.

\subsection{Optimal UAV-swarm path planning} \label{sec:multiple-drone-dp}

We have solved the single UAV problem by determining the UAV's path planning for servicing many locations in Algorithm~\ref{alg:one-general}.
In this subsection, we extend the dimensionality-reduced dynamic programming approach in Algorithm \ref{alg:one-general} to optimally solve UAV-swarm cooperation problem.
The major difference is that here we encounter the situation that multiple UAVs could visit the same location and the same demand could possibly be targeted by multiple UAVs simultaneously.
To determine each UAV's path planning, we record the visiting information of locations and times from all the other $|K|-1$ UAVs.
Moreover, to coordinate the UAV routing actions, we focus on a time interval starting from time $t$ (e.g., time interval $(t,t+t_k^{*})$ below) for each UAV during which the UAV sticks to the last routing action (i.e., either staying at the current location or on the way traveling to some target location).
Different from the approach in Section~\ref{sec:single-drone}, we encode the decision state at time point $t$ by a long tuple $\pe{Q,i, t, \hat{s}_1,\hat{s}_2,...,\hat{s}_{|K|},{t}^{*}_1,{t}^{*}_2,...,{t}^{*}_{|K|}}$, which  tells that
\begin{itemize}
\item
a number of $i$ demands are serviced up to time $t$,
\item
UAV $k \in K$ will be at location $\hat{s}_k$ at future time $t + {t}^{*}_k$,
\item
and each tuple $\pe{s',t'} \in Q$ indicates that a UAV leaves location $s' \in S$ at historical time $t'$.
\end{itemize}

\noindent
In the decision state described above, we follow the principle that each UAV $k$ is on the way to the target location $\hat{s}_k$ during time interval $(t,t+ {t}^{*}_k)$.
This can be interpreted as two cases:  either the UAV is flying to the target location or it is hovering at the current location for the next demand to be released at time point $t+ {t}^{*}_k$.
For both cases, new decisions have to be made at time point $t+ {t}^{*}_k$.
%
%$\pe{Q,i, t, \hat{s}_1,\hat{s}_2,\ldots,\hat{s}_{|K|},{t}^{*}_1,{t}^{*}_2,\ldots,{t}^{*}_{|K|}}$
%$t + {t}^{*}_{\hat{k}}$.
Both lemmas~\ref{lmm:prnc-1} and \ref{lmm:prnc-2} are still useful here to simplify the routing decisions.
Similar to Algorithm~\ref{alg:one-general}, set $Q$ about the UAVs' departure time stamps at all the locations can be encoded with complexity $O(n^{(\alpha-1) |K|} |S|^{(\alpha-1) |K|} )$, which now further depends on UAV number $|K|$ and thus we extend Algorithm~\ref{alg:one-general} by using the new decision state for the UAV-swarm.
%Moreover, the computational complexity for each state depends on time ${t}^{*}_{\hat{k}}$, which can be bounded by the maximum pairwise travelling distance $a_{max}$.

\begin{lemma}\label{lmm:multi-dp}
The optimal solution of the UAV-swarm's cooperative path planning can be obtained with high computation complexity 
$O(n^{1 + (\alpha-1) |K|} |S|^{1 + \alpha |K|}  a_{max}^{|K|} )$ 
for a large number $|K|$ of UAVs, where $a_{max}$ is the longest traveling time between any two locations in set $S$.
\end{lemma}
%\begin{proof}

\noindent
\textit{Sketch of proof.} Similar to variable $g(Q,i,s) = t$ used in Algorithm~\ref{alg:one-general} to represent decision state  $\pe{Q,i,s,t}$ for a single UAV, here we use variable $g({Q,i,\hat{s}_1,\hat{s}_2,\ldots,\hat{s}_{|K|},{t}^{*}_1,{t}^{*}_2,\ldots,{t}^{*}_{|K|}}) = t$ to represent decision state $\pe{Q,i, t, \hat{s}_1,\hat{s}_2,\ldots,\hat{s}_{|K|},{t}^{*}_1,{t}^{*}_2,\ldots,{t}^{*}_{|K|}}$ for $|K|$ UAVs.
Algorithm~\ref{alg:one-general} can thus be generalized to compute the new variables here. The optimal path planning for multiple UAVs can be computed as long as all possible states and all possible routing decisions are considered in the computation.

To extend the state $\pe{Q,i, t, \hat{s}_1,\hat{s}_2,...,\hat{s}_{|K|},{t}^{*}_1,{t}^{*}_2,...,{t}^{*}_{|K|}}$ and generate new partial path planning, we aim to consider the routing decision at the earliest time that a UAV arrives its target location.
Specifically, we find the earliest time $t' = t + {t}^{*}_{\hat{k}}$ when some UAV $\hat{k}$ arrives its target location $\hat{s}_k$.
During time interval $(t,t')$, all the UAVs are on their ways to their target locations, and hence we only need to focus on the routing decision at time point $t'$.
We enumerate all the possible locations for the UAV $\hat{k}$ to visit after time $t'$, i.e., either continue staying at the current location $\hat{s}_k$ or flying to some other location.
For each of such options, we update the variables $g(\cdot)$ for the new states, using the similar approach as in Algorithm~\ref{alg:one-general}.
As a result, we prove that  Algorithm~\ref{alg:one-general} can be generalized to compute the optimal path planning.

Now we analyze the computational complexity.
Generalizing to multiple UAVs, set $Q$ can be encoded with complexity $O(n^{(\alpha-1) |K|} |S|^{(\alpha-1) |K|} )$.
The complexity for variables $g(\cdot)$ now becomes $O(Q) \cdot O(n |S|^{|K|} a_{max}^{|K|})$. Additionally, $O(|S|)$ is needed to compute each variable $g(\cdot)$.
Overall, the computational complexity is
$O(n^{1 + (\alpha-1) |K|} |S|^{1 + \alpha |K|}  a_{max}^{|K|} )$.
%\end{proof}

We can see from Lemma \ref{lmm:multi-dp} that the refined Algorithm~\ref{alg:one-general} returns the optimal solution but with formidably high computational complexity.
This motivates us to propose new low-complexity algorithms for a large number of UAVs to cooperate.

% $O(n^{\alpha |K| + 1} |S|^{\alpha |K| + |K|}  a_{max}^{|K|})$.

\subsection{Benchmark: partition-based UAV-swarm routing algorithm} \label{subsec:benchmark}
It is straightforward to decompose the complex problem and partition the UAV-swarm into $|K|$ subproblems, by assigning the location clusters to UAVs.
%by determining each UAV's path planning separately.
In this benchmark case, we divide all the locations into different subsets/clusters and assign one UAV to service each subset of locations. We call this benchmark approach as {\em partition-based UAV-swarm routing algorithm}, where any two UAVs will not visit the same location.
\begin{figure}[!ht]
%\centerline{\includegraphics[width = 0.5 \columnwidth]{figsplit.pdf}}
\centering
\subfloat[Trajectories at optimum]{
\begin{minipage}[t]{0.47\linewidth}
\centering
\includegraphics[keepaspectratio, height = 40mm]
{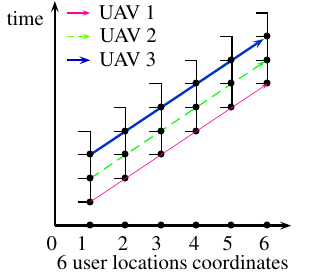}
\label{Fig.split.a}
\end{minipage}
}
\centering
\subfloat[Trajectories from partition-based algorithm]{
\begin{minipage}[t]{0.47\linewidth}
\centering
\includegraphics[keepaspectratio, height = 40mm]
{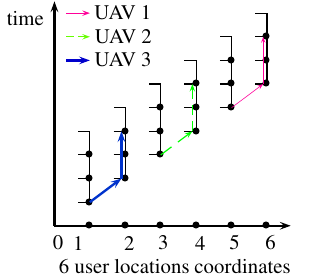}
\label{Fig.split.b}
\end{minipage}
}
\caption{The trajectories from the optimal algorithm and the best partition-based algorithm are depicted in Fig.~\ref{fig:partition}\protect\subref{Fig.split.a} and Fig.~\ref{fig:partition}\protect\subref{Fig.split.b} respectively, where the three lines with red, green, blue colors indicate the trajectories of the three UAVs, respectively.
At the optimum, the UAVs visit sequentially each location and all demands are serviced, while only $12$ demands are serviced in partition-based algorithm.
%However, when each UAV is assigned to cover two locations independently, as in Figure~\protect\subref{Fig.split.b}, only four demands can be serviced by each UAV, resulting approximation ratio $12/18 = 0.667$.
}
\label{fig:partition}
\end{figure}
%However, if the $|S|=6$ locations are partitioned among the $|K|=3$ UAVs ($2$ locations for each), as in Figure~\protect\subref{Fig.split.b}, each UAV only service four demands. The approximation ratio is thus $12/18 = \frac{1}{|K|} + \frac{|K|-1}{|S|} = 0.667$.
%In Figure~\protect\subref{Fig.split.b}, each UAV is assigned to two locations and successfully services four demands.
%\noindent
We show that any partition-based UAV swarm-routing algorithm has approximation ratio no better than $\frac{1}{|K|} + \frac{|K|-1}{|S|}$ in the worst case, and Fig.~\ref{fig:partition} illustrates the big gap between the optimal solution and the partitioned-based algorithm. 
%
%PLACE FIG. 3 HERE ON THE SAME PAGE AS FIRST MENTIONED 
%as given in Lemma~\ref{lmm:nonshare}.
%
%Lemma \ref{lmm:nonshare}.% is $\frac{1}{|K|} + \frac{|K|-1}{|S|}$.

\begin{lemma}\label{lmm:nonshare}
%Given $|K| < |S|$, 
There exists an instance such that any partition-based UAV-swarm routing algorithm has approximation ratio no more than $\frac{1}{|K|} + \frac{|K|-1}{|S|}$ in the worst case.
When the UAV number $|K|$ and the location number $|S|$ go to infinity, this ratio $\frac{1}{|K|} + \frac{|K|-1}{|S|}$ approaches zero.
\end{lemma}
\begin{proof}
Consider the worst-case instance where all the locations are on a straight line and the traveling time between any two adjacent locations is one and demand service time is arbitrarily small ($q \rightarrow 0$).
Assume the $|S|$ locations are ordered from leftmost to rightmost.
For each location $s \in \se{1,2,...,|S|}$, we create a number of $|K|$ demands of unit waiting time where the release times are $\se{s+0,s+1,...,s+|K|-1}$.
The instance for $|S| = 6, |K| = 3$ is depicted in Fig.~\ref{fig:partition}.

In an optimal solution in Fig.~\ref{fig:partition}\protect \subref{Fig.split.a}, each UAV will start from the leftmost location with starting times $\se{0,1,...,|K|-1}$, and fly to the next location (i.e., from location $s$ to $s+1$) immediately after servicing one demand, thus the first UAV will finish the earliest released demand on each location, the second UAV will finish the second earliest released demand on each location, and so on. Therefore, all demands are serviced in the optimal solution.

For partition-based solutions, we claim that in any feasible partition-based solution (e.g., Fig.~\ref{fig:partition}\protect \subref{Fig.split.b}), any UAV that visits a number of $k'$ distinct locations will finish no more than $k'+|K|-1$ demands.
The claim will result in the conclusion that at most $\sum_{k \in K} (i_k + |K|-1)$ demands are serviced in the best partition-based solution with $i_k$ being the number of locations assigned to the $k$-th UAV, which is bounded by $|S| + |K|(|K|-1)$ due to the fact that each location is only allowed to be visited by at most one UAV.

Now we prove the claim.
A location is {\em visited} by a UAV if the UAV finishes at least one demand at the location.
Consider a UAV in a feasible partition-based solution, and suppose that the UAV has visited a number of $k'$ distinct locations where $s,s' \in \se{1,2,...,|S|}$ with $s \le s'$ are the two special locations with the smallest and largest index, respectively. Consider the demands on these $k'$ locations, the earliest demand is released at location $s$ at time point $s+0$, and the latest demand is released at location $s'$ at time point $s' + |K|-1$. Hence, the UAV can finish demands during at most $t_1 = (s' + |K|) - (s+0)$ units of time because during each unit time there is at most one demand released at each location.
Traveling from location $s$ to $s'$ will waste at least $t_2 = (s'-s)-(k'-1)$ units of time where at each of such unit times the UAV does not service any demand, i.e., the UAV flies over some location without staying at that location.
%Moreover, at each unit time, the UAV can finish at most one demand.
Therefore, the UAV finishes at most $t_1 - t_2 = k' + |K| - 1$ demands and the claim holds.
As a consequence, the approximation ratio of any partition-based solution is no better than  $\frac{|S| + |K|(|K|-1)}{|S| |K|} = \frac{1}{|K|} + \frac{|K|-1}{|S|}$.
\end{proof}
%\fi
%For large $|K|$ and large $|S|$, this ratio $\frac{1}{|K|} + \frac{|K|-1}{|S|}$ approaches to zero. Hence, the best partition based solution can be arbitrarily bad.

\subsection{Fast iterative algorithm for UAV-swarm path planning}\label{subsec:iterative-greedy}

%Our main focus is the first method, as a comparison we show that %where we  We develop the first method  ,
%We propose a greedy algorithm for multiple UAVs in this section. To be clear, with respect to the single UAV problem, there could also exist greedy algorithm with good approximation, which differs from our greedy approach.
%Before we present our greedy algorithm, we show that for multiple UAVs, without sharing locations the solutions can be arbitrarily bad in the worst case scenario. We describe the result in the following paragraph.
%
%the dynamic program in Section \ref{sec:multiple-drone} is too complex and
We have shown that the partition-based UAV-swarm routing algorithm without location overlap in Section \ref{subsec:benchmark} can have arbitrarily poor  performance in the worst case, though it has low complexity.
Alternatively, here we present a fast iterative algorithm by iteratively routing each UAV one by one
%sending UAVs one by one
with location overlap for the UAV-swarm cooperation. 
In each iteration, we compute the best path planning solution for routing an individual UAV based on Algorithm~\ref{alg:one-general} with the input of the demands that have not been serviced by any UAV yet, and finally remove the already serviced demands by that UAV.
%we record the set of missed demands,
We continue this process until all the demands are serviced or all the UAVs are used up. We call this  algorithm as {\em Fast Iterative UAV-swarm Routing Algorithm}, which is presented in Algorithm~\ref{alg:greedy} by calling Algorithm~\ref{alg:one-general} in Line \ref{alg-line-callsingle}.

More specifically, in the process of running Algorithm~\ref{alg:greedy}, let $J_u$ be the set of demands that have not been serviced by any UAV yet.
For each UAV $k \in K$, we obtain demand set $J(k) \subseteq J_u$ by running Algorithm~\ref{alg:one-general} such that all the demands from $J(k)$ can be serviced by this single UAV.
%one more UAV and the immediate hit of user demands is maximized.
Afterwards, we remove demands $J(k)$ from set $J_u$, i.e., $J_u \gets J_u \setminus J(k)$ in Line~\ref{alg-line-greedyreduce} of  Algorithm~\ref{alg:greedy}.

%Providing that each UAV's trajectory can be obtained, i.e., either an optimal or approximated solution can be found.

\begin{algorithm}[!tbh]\small
%\footnotesize
\caption{Fast Iterative UAV-swarm Routing Algorithm.}
\label{alg:greedy}
\begin{algorithmic}[1]
\State $J_u \gets J$
\For{$k \in \se{1,2,...,|K|}$}
\State call Algorithm~\ref{alg:one-general}
for individual UAV $k$ to service demand set $J_u$
\label{alg-line-callsingle}
\State $J(k) \gets $ serviced demands by UAV $k$
\State $J_u \gets J_u \setminus J(k)$
\label{alg-line-greedyreduce}
\EndFor
\State {\bf Return} demands $\cup_{k \in K} J(k)$.
\end{algorithmic}
\end{algorithm}

%\begin{figure}[!t]%[htbp]
%\centerline{
%\includegraphics[width = 0.9 \columnwidth]{figure_greedy.pdf}
%}
%\caption[12pt]{Comparison between approximation ratio of Fast Iterative UAV Swarm Cooperation Algorithm and the upper bound of partition based UAV-swarm routing algorithm as the increase of the number of UAVs. The black line indicates the approximation ratio for our Fast Iterative UAV Swarm Algorithm, and the other six lines indicates the upper bound for the partition based solutions with respect to the number of locations $|S| = 6,9,12,15,18,21$.}
%\label{fig-greedy}
%\end{figure}

%Fast Iterative Algorithm for UAV Cooperation
\begin{proposition}\label{lmm-greedy-cover}
Algorithm~\ref{alg:greedy} nicely achieves constant approximation ratio $1 - (1-1/|K|)^{|K|}$ in the worst case, which is greater than $1-1/e$.
\end{proposition}

%The proof technique of the approximation ratio is similar to \cite{hochbaum1998analysis} for the analysis of maximum $k$-coverage problem.

%\iffalse
\begin{proof}
The proof idea of the approximation ratio is inspired by \cite{hochbaum1998analysis} for the analysis of maximum $k$-coverage problem.
In this proof, we write $w(J^{\#}) = \sum_{j \in J^{\#} } 1$ as the total number of demands in $J^{\#} \subseteq J$.
For each UAV $k \in K$, we denote
$J(k), \tilde{J}(k)$ as the sets of demands which are serviced by UAV $k$ in Algorithm~\ref{alg:greedy} and the optimum, respectively.
Especially, let $J^{*} = \cup_{k \in K} \tilde{J}(k)$ be the set of serviced demands in the optimal solution.

Next, we prove for each $k \in \se{1,2,\ldots,|K|}$, it holds that
\begin{equation}\label{eqn-greedy}
w( J(k) ) \ge \frac{1}{|K|} \cdot w (J')
\end{equation}
\noindent where $J' = J^{*} \setminus (\cup_{k'=1}^{k-1} J(k'))$.
When UAV $k$ is considered in Algorithm~\ref{alg:greedy}, demand set $J'$ indicates the set of demands which are not serviced yet but instead serviced by the optimum.
By the pigeonhole principle, there exists UAV $\hat{k} \in K$ in the optimal solution such that $ w( J' \cap \tilde{J}(\hat{k}) ) \ge \frac{1}{|K|} \cdot w( J' )$.
In other words, at least a portion of $\frac{1}{|K|}$ demands from $J'$ are serviced by some UAV alone at optimum, 
%at least one UAV services a portion of $\frac{1}{|K|}$ demands from $J'$, 
which is true because all demands from $J'$ are serviced in optimum.
Therefore, in Algorithm~\ref{alg:greedy} UAV $k$ can at least follow the same solution of UAV $\hat{k}$ in the optimal solution.
Because Algorithm~\ref{alg:one-general} is optimal for a single UAV, it implies that the total number of demands serviced by UAV $k$ in Algorithm~\ref{alg:greedy} is at least that of
UAV $\hat{k}$ in the optimal solution, i.e., $ w( J' \cap \tilde{J}(\hat{k}) )$.
That is, $w ( J(k) ) \ge w ( J' \cap \tilde{J}(\hat{k}) ) \ge \frac{1}{|K|} \cdot w (J')$.

Now, we are ready to prove the approximation ratio.
We denote optimum as $OPT = w (J^{*})$, and
let $W(k) = w ( \cup_{k'=1}^{k} J(k') )$ be the total number of serviced demands by the UAVs up to UAV $k$ iteration in Algorithm~\ref{alg:greedy}, especially let $W(0) = 0$.
Because one demand is serviced by at most one UAV in our algorithm, Eq.~\eqref{eqn-greedy} can be equivalently expressed as
\begin{equation}
W(k) - W(k-1) \ge \frac{1}{|K|} \cdot (OPT - W(k-1)), \notag
\end{equation}
\noindent as $W(k) - W(k-1) = w(J(k))$ and $w(J') = w(J^{*}) - w(\cup_{k'=1}^{k-1} J(k')) = OPT - W(k-1)$.
% for all $k \in \se{1,2,...,|K|}$ , which is
Then reorganizing the equation, we have
\begin{equation}
W(k) - OPT \ge (1-\frac{1}{|K|}) \cdot (W(k-1) - OPT). \notag
\end{equation}
Proceeding by induction from $k = |K|$ to $k = 1$, we have
\begin{equation}
W(|K|) - OPT \ge (1-\frac{1}{|K|})^{|K|} \cdot (W(0) - OPT). \notag
\end{equation}
That is,
\begin{equation}
W(|K|) \ge ( 1 - (1-\frac{1}{|K|})^{|K|} ) \cdot OPT. \notag
\end{equation}
Note that $W(|K|)$ indicates the total number of demands serviced in  Algorithm~\ref{alg:greedy}, hence
Algorithm~\ref{alg:greedy} has constant approximation ratio $1 - (1-\frac{1}{|K|})^{|K|}$.
\end{proof}
%\fi

Note that the approximation ratio $ 1 - (1-\frac{1}{|K|})^{|K|}$ decreasingly approaches to $1-1/e$ when $|K|$ approaches to infinity, where $e \approx 2.718$ is the base of the natural logarithm. Algorithm~\ref{alg:greedy} performs much better than the partition-based algorithm with zero ratio in the worst-case, especially for a large number of UAVs $|K|$ and a large number of locations $|S|$ as shown in Lemma~\ref{lmm:nonshare}. 
Later in the simulation experiment section, we will show that the average-case performance of Algorithm~\ref{alg:greedy} is close to optimum.
Moreover, Algorithm~\ref{alg:greedy} has low computational complexity $O(|K| n^{\alpha} |S|^{\alpha+1})$, which is linear to the number of UAVs $|K|$ and the term $O(n^{\alpha} |S|^{\alpha+1})$ is the complexity for solving the optimal path planning of a single UAV as shown in Proposition~\ref{prop:time-one-drone}.

%There exists a path planning in which UAVs do not collide, and 
\begin{proposition}\label{lmm-greedy-collision}
There exists a collision-free path planning that the total number of successfully serviced demands is the same as that of Algorithm~\ref{alg:greedy}.
\end{proposition}
\begin{proof}
We identify two cases when two UAVs collide, i) they service the same location at the same time, ii) they meet during traveling to its target location. 
When any of these two cases happens, we show that the routing path of the UAVs returned by Algorithm~\ref{alg:greedy} can be transformed to avoid UAV collision, without changing the serviced demands. Specifically, whenever two UAVs meet, we swap their future paths and adjust the new paths to avoid collision.

For case i), we swap the departure times of the two UAVs at this location, and thus the time period servicing the location by one UAV will be contained by that of the other UAV, resulting in the fact that the demands serviced by one UAV can actually all be serviced by the other UAV.
Therefore, we adjust its path to skip visiting this location and hence avoid the collision.

For case ii), we swap the target locations of the two UAVs when they meet, resulting in the paths that the two UAVs will return back to its departure location, which is a waste of time.
Therefore, we adjust the paths to skip such meaningless turnarounds, and hence avoid UAV collision.
\end{proof}

\section{Algorithms for UAV-Swarm Powered with Battery Charging Stations}\label{sec:energy}

In previous sections, we have successfully designed and analyzed cooperative path planning algorithms, where each UAV has a finite time in the sky to route and service due to the limited battery capacity. 
%under UAVs' limited battery capacity. 
In this section, to provide sustainable service provisioning, we further consider that each UAV can charge its battery by reaching an battery charging 
station. Accordingly, we jointly design UAVs' cooperative path planning over not only users' locations but also battery charging stations in the long run. 
The battery charging option introduces new decision choices into UAV path planning and makes it more challenging to design the cooperative path planning algorithm.
In the following, we first extend the system model with battery charging and show %Lemma~\ref{lmm:prnc-1}, Lemma~\ref{lmm:prnc-2} and 
%characterize the major difference of the optimal solution when battery charging is involved.
%We find that directly applying our proposed algorithms here will not give optimal solution since the 
that the dominance property of optimal path planning in Lemma~\ref{lmm:dominated} do not hold anymore. 
We then analyze new dominance property and refine our previous algorithms.
%with similar performance.
%The design becomes more challenging and multiple routing paths will be required for UAVs' spatial-temporal information of charging. 
%
%In this section, we generalize the problem into a more practical case  with the consideration of UAV battery  charging.

On top of the system model defined in Section~\ref{sec:problem-description}, 
we additionally consider a set $C$ of battery charging stations distributed spatially on the ground to power each UAV with fixed battery capacity $B$.
We assume each UAV has power consumption rate $p_0$ per unit time when hovering and $p_1$ when flying.
Without loss of generality, we assume $p_0 \le p_1$.
%with two flying mode \emph{hovering} and \emph{flying}, with power consumption rate $p_{0}$ and $p_{1}$ respectively. 
With remaining energy $B_t$ in battery, it takes time $t_c(B_t)$ for a UAV to fully charge the battery to capacity $B$.
%, where time $t_c(B_t)$ can be a constant if UAV battery is replaced instead of charged.
% We generalize the travelling distance matrix defined in Section~\ref{sec:problem-description} to include the charging stations $C$ with similar triangle inequality in Equation~\eqref{eq:triangle}.
%To simplified the algorithm design, we denote $B_s^{*}$ as the minimum amount of energy for travelling from  location $s \in S$ to the nearest charging station from $C$, which can be calculated as follows.
%\begin{equation}\label{eq:min-energy}
%    B_s^{*} = \min_{c \in C} \limits ~ p_1 \cdot a(s,c)
%\end{equation}
We require the path planning that each UAV starts at some charging station and finally lands at a charging station when finishing servicing the assigned demands.
%To simplify the analysis, we do not allow partially charging options, i.e., the UAV battery can only be charged to full capacity or not charged at all. Moreover, when the charging process completes, we allow the UAV to wait on the ground for some extra time for new demands to be released.
Yet we flexibly allow a UAV (after fully charged) to wait extra time on the ground charging station for new demands to be released.

Based on the system model in Section~\ref{sec:problem-description}, we update the UAV  routing constraint in Eq.~\eqref{eq:obj} in the following equation. With charging, for each UAV $k$, the set $T_{k,s}$ includes the time periods when the UAV stays at location $s$, where $s \in \se{1,2,...,|S|}$ indicates a service location, and $s \in \se{|S|+1,|S|+2,...,|S|+|C|}$ indicates a charging location.
The routing defined by $T_{k,s}$ is feasible if the battery level is non-negative at each time points of  $T_{k,s}$.
\begin{equation}
(T_{k,1},T_{k,2},\ldots,T_{k,|S|+|C|}) \longrightarrow \text{feasible routing with charging}
\end{equation}

With such model extension to include charging stations, we first refine our dominance property in  Lemma~\ref{lmm:dominated} to fit the new charging model.
%
%we first refine our lemmas~\ref{lmm:prnc-1} and \ref{lmm:prnc-2} to fit the new model. 
Our previous algorithms in Sections~\ref{sec:single-drone} and \ref{sec:multiple-drone} can be regarded as a special case here for only one battery cycle of UAV operation before using up energy storage $B$. 
%That is to say, they aim to service the maximum number of demands up to a fixed time $T$, which is the maximum time that a UAV can stay in the air by support of one battery cycle.
With the consideration of battery charging, however, the problem becomes complex since each UAV needs to decide when and where (not necessarily the nearest charging station) to charge the battery.
Considering its energy consumption, it is no longer optimal for each UAV to service the demands as early as possible, which violates the dominance property in Definition~\ref{def:dominanted}.
%This can be seen from the intuition that the UAV has to wait in the air with a longer time for the next demand to be released when it finishes servicing the previous demands earlier and hence consumes more energy during waiting.
%with the support of flexible waiting time on the ground 

This can be seen from an example that after charging, the UAV can indeed return back to some user location as early as possible or 
exactly at the time when a new demand is released there.
However, this may not be optimal since after servicing this demand, the UAV may have to consume more energy to hover in the air for a longer time for the next demand to be released. In this situation, a better strategy is just to wait on the ground for more time after charging such that this demand is serviced with the later released demands together.
%
%An example is that the UAV needs to hover for a longer time to wait for the new demands to be released if it finishes servicing the last demand earlier. 
%On the other hand, arriving at the user location late will miss the demands.
%
%
As a result, the UAV now needs to reach the user location within the 
overlapping
%``overlapping''
waiting time window of the user demands as much as possible (i.e., neither earlier nor later).
%As a result, the UAV now needs to arrive at the latest at the user location to just not miss the target demand there.
%%the routing algorithm now needs to send the UAV to the user location as much as possible within the waiting time window of the user demand.
%
Except for this, both lemmas \ref{lmm:prnc-1} and \ref{lmm:prnc-2} still work here for characterizing the UAV movement in the optimal solution, since any violation of these two lemmas will lead to unnecessary waste of time or energy.
Based on these observations, we still manage to characterize the decision states of the MDP problem.
%, as well as the times for making routing decisions (including charging decisions).

Inspired by the design idea of Algorithm~\ref{alg:one-general} in section~\ref{subsec:dp-transition}, we first propose a dimensionality-reduced dynamic programming based algorithm to obtain the optimal path planning of a single UAV with battery charging.
Similar to Algorithm~\ref{alg:greedy} in Section~\ref{subsec:iterative-greedy}, for the whole UAV-swarm problem with battery charging, we then 
%Then, we propose an integer linear programming (ILP) formulation to compute the optimal path planning for the UAV-swarm. The ILP formulation is based on the feasible path planning for a single UAV. In our previous dynamic programming approach, each feasible path planning of a single UAV corresponds to a path on the directed acyclic graph (DAG) of the state transition diagram, we use this DAG to formulate the path planning for all UAVs by integer linear programming approach. Though ILP can solve the problem optimally, it may consume unacceptable computational time. 
present a fast iterative greedy algorithm with low complexity and prove its theoretical performance.

%Algorithm~\ref{alg:greedy} to obtain the path planning for multiple UAVs with the same provable approximation ratio (new algorithm is shown in Algorithm~\ref{alg:greedy-charging}).

%In the following, we first generalize the dynamic programming algorithm to obtain the optimal path planning of a single UAV with battery charging.
%Based on this, we show that our proposed greedy algorithm (Algorithm~\ref{alg:greedy}) can also be applied to obtain the path planning for multiple UAVs, with the same provable approximation ratio.
%Additionally, we propose an integer linear programming (ILP) formulation to compute the optimal path planning for the UAV-swarm.
%The ILP formulation is based on the feasible path planning for a single UAV.
%In our previous dynamic programming approach, each feasible path planning of a single UAV corresponds to a  path on the directed acyclic graph (DAG) of the state transition diagram, we use this DAG to formulate the path planning for all UAVs by integer linear programming approach.

\subsection{Optimal path planning of a single UAV with battery charging}\label{subsec:energy-state}
In this subsection, we characterize the decision states for computing the optimal path planning of a single UAV powered by battery charging stations.
Recall that in Algorithm~\ref{alg:one-general} 
we use set $Q$ to label the UAV departure times at all the locations, where each tuple $\pe{s^{*},t^{*}} \in Q$ 
tells the last UAV departure time $t^{*}$ at location $s^{*}$.
%tells that time $t^{*}$ is the last time when the UAV leaves location $s^{*}$.
At time $t$, the decision state can be represented by a tuple $\pe{Q, B_t, i, s, t}$, indicating that the UAV at location $s$ just services an accumulated number $i$ of demands by time $t$ with remaining battery energy $B_t$ and latest departure times $Q$.
%at locations. 
Compared with the decision state in Section~\ref{sec:single-drone}, we only need to include the energy storage state $B_t$ in the decision state here.
However, we cannot directly adopt the variable $g(Q,i,s)$ from Algorithm~\ref{alg:one-general} to compute the optimal path planning because servicing demands at the earliest may not be the optimal strategy.
%path planning strategy
%abuse notation $g(\cdot)$ and
Instead, as the UAV prefers to finish more demands with more energy left in battery, decision state $\pe{Q, B_t, i, s, t}$ dominates $\pe{Q, B_t', i, s, t}$ if $B_t > B_t'$.
%when comparing different decision choices, 
Therefore, we redefine $g(Q,i,s,t)$ as the maximum energy that the UAV at location $s$ can keep in battery at time $t$ for servicing an accumulated number of $i$ demands under the constraints of departure times $Q$.
%
%In other words, 
Since the dominance property on time domain does not hold, now the computational complexity of variables $g(Q,i,s,t)$ relies on the value of time $t$. %We still make the restriction for the time of decisions such that a demand is finished by the UAV exactly at time $t$.

%In term of state transition, we need to consider the new charging decision when the UAV decides to leave the current location $s \in S$ to another location $s' \in S$. 
%That is to say, the UAV can choose to charge the battery at some ground station before flying to the target location $s'$.
In term of charging decisions, the UAV now can choose to charge the battery at some ground station $c \in C$ before flying from the current location $s \in S$ to some target location $s' \in S$.
%to the target location $s'$. we need to consider the new charging decision when the UAV decides to leave the current location $s \in S$ to another location $s' \in S$. 
We describe all possible subsequent decisions based on the current state.
%
%Suppose $g(Q,i,s,t) = B_t$ is computed, i.e., the UAV routing decision is made based on state $\pe{Q, B_t, i, s, t}$.
%a demand is finished at time $t$ at location $s$. 
%
%
%Suppose the UAV reaches decision state  $\pe{Q, B_t, i, s, t}$, similar as in Algorithm~\ref{alg:one-general},  we  restrict the state such that a demand is finished by the UAV exactly at time $t$ and regard time $t-q$ as the decision-making time stamps due to service time $q$.
%In the new algorithm, we enumerate all possible subsequent decisions after time $t-q$ and identify the corresponding 
In new decision state $\pe{Q', B_{t'}, i', s', t'}$, we disregard the choices such that the resulting battery energy $B_{t'}$ is not sufficient for the UAV to reach the nearest charging station after time $t'$, i.e., $B_{t'} < p_1 \cdot \min_{c \in C} a(s',c)$.
Similar to Algorithm~\ref{alg:one-general} in Section~\ref{subsec:dp-transition}, we consider time $t-q$ as a decision-making time stamp only if some demand will be serviced at time $t$ as it takes time $q$ to service a demand.
%Note that it takes time $q$ to service a demand, the UAV leaving from location $s$ at time $t$ decides to stay or leave at time $t-q$.
We separate the state transition analysis associated with the possible decisions at time $t-q$ into the following two cases.
\begin{itemize}
%\item If the UAV decides at time $t-q$ to stay at location $s$ to service more demands, the next demand release time $t^{*}$ can be calculated as in  Algorithm~\ref{alg:one-general}. Then, at time $t' = t^{*} + q$ the UAV will finish servicing this new demand with the remaining energy 
\item If the UAV decides to stay at location $s$ at time $t-q$ to service more demands, the earliest time for the next demand to be released at location $s$ is $t^{*} = \min \se{r_j ~|~ t - q < r_j, j \in J_s}$. The UAV will finish servicing this demand at time $t^{*} + q$. In this case, we have $Q' = Q, s' = s, t' = t^{*} + q$, $i' = i + \sum_{j \in J_{s'}: r_j = t^{*}} 1$ and the energy level $B_{t'}$ at time $t'$ is updated to:
\begin{equation}
    B_{t'} = B_t - p_0 (t' - t).
\end{equation}

% at location $s$
\item Otherwise, if the UAV decides to leave at time $t-q$, it does not service any new demand during time interval $(t-q,t]$. After time $t$, the UAV can fly to the target location $s' \in S$ with or without battery charging on the way, as discussed in the following two sub-cases. 
%We separate this decision into two sub-cases, depending on whether the UAV energy is charged or not.

\begin{itemize}
\item The UAV directly flies to new location $s' \in S$ without battery charging, i.e., it will reach location $s'$ at future time $t^{*} = t + a(s,s')$.
By Lemma \ref{lmm:prnc-2}, the UAV should at least service one demand at location $s'$ and let $t'- q$ (with $t' - q \ge t^{*}$) be the earliest time to  service a new demand.
Similar to Algorithm~\ref{alg:one-general}, 
time $t'$ can be calculated based on the previous departure times in $Q$ and the demand information at location $s'$.
For the remaining energy level at time $t'$, it is updated to:
\begin{equation}
B_{t'} = B_t - p_1 \cdot a(s,s') - p_0 (t'-t^{*}).
\end{equation}

\item The UAV decides to charge the battery before flying to location $s'$. 
In this process, the UAV flies to charging station $c \in C$ and possibly another charging station $c' \in C$ later to charge the battery to full capacity.
This is possible if it consumes a lot energy when flying to location $s'$ directly, and it is better to refill from another charging station $c'$ located around location $s'$. 
It is also possible that the UAV stays at charging station $s'$ after fully charged and return to sky at time $t^*$. 
In the new algorithm, we examine every feasible possibility of the charging process, i.e., values for $s',c,c',t*$.
At time $t^{*} + a(c',s')$, the UAV will arrive at the new location $s'$ to service a new demand.
The UAV's battery energy 
%at time $t' = t_5 + q$ can be computed as follows.
after servicing this demand at time $t' = t^* + a(c',s') + q$ is updated to:

\begin{equation}\label{eq:double-charge}
B_{t'} = B - p_1 \cdot a(c', s') - p_0 \cdot q.
\end{equation}

%The reason to consider two charging stations $c,c'$ here is that from location $s$, the UAV might only has enough energy to reach the nearest charging station $c$, while there may be many demands going to be released at the user location $s'$ near the charging station $c'$. Directly flying from charging station $c$ to the user location $s'$ might already consume a lot of energy and lead to less time for servicing the demands there. Hence, it might be better for the UAV to directly fly to charging station $c'$ to charge the battery first and then fly to the nearby location $s'$ to service the demands.
\end{itemize}
\end{itemize}

\noindent
For both cases above, the UAV can finish  servicing a demand at future time $t'$ at location $s'$. We count the additional demands that can be started to service at time $t'-q$ and update parameter $i'$. In the revised procedure above, we successfully refine Algorithm~\ref{alg:one-general} for returning an optimal path planning solution for the single UAV powered by ground charging stations.

%and $T_w \le T$ be the maximum allowed waiting time slot number on the ground
\begin{proposition}
Let $T = \max_{j \in J} d_j - \min_{j \in J} r_j$ be the total time slot number, the variables $g(Q,i,s,t)$ in the revised MDP can be computed in complexity $O(T^2 \cdot n^{\alpha} \cdot |C|^2 \cdot |S|^{\alpha+1})$.
\end{proposition}
% T^2 -> T T_w,
%\begin{proof}
\noindent
\textit{Sketch of proof.} 
Similar to the computational complexity analysis of Algorithm~\ref{alg:one-general} in the proof of  Proposition~\ref{prop:time-one-drone}, 
after servicing this demand at time $t' = t^* + a(c',s') + q$, the computational complexity for 
encoding $g(Q,i,s,t)$ here is $O(Q)\cdot O(n |S| T)$ with $O(Q) = O( n^{\alpha-1} |S|^{\alpha-1})$ being the computational complexity for set $Q$.
It takes $O(T |S| \cdot |C|^2)$ operations to enumerate all possible decisions for joint routing and charging, especially for the last sub-case in Eq.~\eqref{eq:double-charge} above.
In that sub-case, enumerating all possible $s', c, c', t^{*}$ takes $O(|S|), O(|C|), O(|C|),O(T)$ operations, respectively.
Hence, the computational complexity for computing all
the variables $g(Q,i,s,t)$ is  $O(T^2 \cdot  n^{\alpha} \cdot |C|^2 \cdot |S|^{\alpha+1})$.
%$O(T^2  n^{\alpha} \cdot |C|^2 \cdot |S|^{\alpha+1})$.
%\end{proof}

\subsection{Cooperative path planning for the UAV-swarm with battery charging}\label{subsec:lp}
In this subsection, we focus on finding the path planning for the UAV-swarm.
%As a matter of fact, the approximation routing algorithm in Section~\ref{subsec:iterative-greedy} still works here since we can solve the path planning optimally for a single UAV.
%
As an extension of the single-UAV case in Section~\ref{subsec:energy-state}, the difficulty of the problem here increases substantially as the number of the UAVs increases. 
Despite of the problem difficulty, we successfully transform the problem to an integer linear program by creating novel directed acyclic graph (DAG) of the UAV-state transition diagram. 
Then we determine the optimal solution with detailed steps, though the overall complexity is still high. 
% given in Appendix~\ref{ap:ilp}
The detailed description of our ILP approach to compute the optimal UAV-swarm with battery charging can be found in Appendix \ref{ap:ilp}.

To further reduce the complexity, 
%It can be observed from Proposition~\ref{lmm-greedy-cover} and Proposition~\ref{lmm:energy-greedy} that locally optimizing each UAV iteratively (Algorithm~\ref{alg:greedy-charging}) will only give approximated solution.
%Therefore, a global optimization approach is essential to find the optimal solution. To tackle the challenge, 
we treat charging stations as special user locations (to go once out of energy) and focus on all the refined feasible path planning choices for each UAV. As the UAVs are identical, any feasible path planning for one UAV is also feasible for another UAV. Inspired by the iterative path planning idea for the UAV-swarm in Section~\ref{sec:multiple-drone} (i.e., Algorithm~\ref{alg:greedy}), we successfully present a fast iterative algorithm with theoretical performance guarantee for the new UAV-swarm problem with battery charging in Algorithm~\ref{alg:greedy-charging}.
%Moreover, we observe that each feasible path planning for a single UAV can be represented by a path on a directed acyclic graph.
%%When merging the path plannings for all UAVs, we focus on reducing the common serviced demands to achieve the best collaboration between UAVs.

\iffalse
In terms of partition-based path planning, we create a binary variable $y_{k,s}$ for each UAV $k \in K$ and each location $s \in S$, where $y_{k,s} = 1$ indicates that location $s$ can only be visited by UAV $k$.
Meanwhile, we add the following constraints into the formulation.

\begin{align} \label{eq:6}
\sum_{k \in K} y_{k,s} \le 1, &~ \forall s \in S
\end{align}

\begin{equation}
\begin{split} \label{eq:7}
  x_{k,e} -  y_{k,s} \le 0, &~ \forall k \in K, \forall s \in S, \\
  &~ \forall v \in V: s(v) = s, \forall e \in \text{IN}(v) \cup \text{OUT}(v)
\end{split}
\end{equation}

Equation~\eqref{eq:6} indicates that each location can only be visited by at most one drones, and Equation~\eqref{eq:7} indicates that if UAV $h$ is not allowed to visit location $s$, then any edge $e$ that contains a node which includes location $s$ cannot occur in the path of UAV $k$.

\fi

\begin{algorithm}[!tbh]\small
%\footnotesize
\caption{UAV-swarm algorithm with battery charging.}
\label{alg:greedy-charging}
\begin{algorithmic}[1]
\State $J_u \gets J$
\For{$k \in \se{1,2,...,|K|}$}
\State {\bf call} refined dynamic programming algorithm in Section~\ref{subsec:energy-state} 
to route UAV $k$ to demand set $J_u$
\State $J(k) \gets $ serviced demands by UAV $k$
\State $J_u \gets J_u \setminus J(k)$
\EndFor
\State {\bf Return} demands $\cup_{k \in K} J(k)$.
\end{algorithmic}
\end{algorithm}

Algorithm~\ref{alg:greedy-charging} iteratively finds the optimal path planning for each individual UAV 
by calling the refined single-UAV algorithm in Section~\ref{subsec:energy-state}.
As it only incurs one UAV at a time and hence can be solved efficiently by our proposed dimensionality-reduced dynamic programming approach. 
%
%show that our previously proposed greedy algorithm (Algorithm~\ref{alg:greedy}) can also be applied here to obtain the path planning for multiple UAVs, which is given in Algorithm~\ref{alg:greedy-charging}.
%
In Algorithm~\ref{alg:greedy-charging}, UAVs are dispatched to spatial locations and charging stations in a greedy manner, as a result, its computational complexity is just linear to the number of UAVs $|K|$, which is given by $O(|K| \cdot T^2 \cdot  n^{\alpha} \cdot |C|^2 \cdot |S|^{\alpha+1})$.
\begin{proposition}\label{lmm:energy-greedy}
%With UAV battery charging, if the optimal path planning of a single UAV can be solved, 
%Algorithm~\ref{alg:greedy} can still be applied to obtain an approximated path planning for multiple UAVs with the same approximation ratio as in Proposition~\ref{lmm-greedy-cover}.
Algorithm~\ref{alg:greedy-charging} achieves constant approximation ratio $1 - (1-1/|K|)^{|K|}$ in the worst case, which is greater than constant $1-1/e$.
\end{proposition}
%\begin{proof}
\textit{Sketch of proof.}
Algorithm~\ref{alg:greedy-charging} works based on the fact that the path planning of one UAV can also be feasibly applied to another UAV since UAVs are identical, i.e., they have the same flying speed and power consumption rate.
Each feasible path planning choice for a UAV can be mapped to a subset of serviced demands by the UAV.
As Algorithm~\ref{alg:greedy-charging} optimizes the path planning for each UAV $k$ individually, it at least uses the same routing strategy as one of the $|K|$ UAVs from the globally optimized solution.
Due to this iterative nature of Algorithm~\ref{alg:greedy-charging}, it can always guarantee that at least a portion of $\frac{1}{|K|}$ demands from the current set $J_u$ of demands are serviced.
As a result, applying mathematical induction, the approximation ratio bound $1 - (1-1/|K|)^{|K|}$ can be calculated at the worst case.
%As long as ... (\KK{TBD})
%The problem can still be related to the maximum k-coverage problem as the same demand is counted as one when it can be serviced by multiple UAVs.
%As a result, the analysis in Proposition~\ref{lmm-greedy-cover} also holds here.
%It is exactly identical to the maximum k-coverage problem except that in our problem it is impossible to enumerate all possible path plannings for a single UAV.
%This can be observed that the optimal path planning for a single UAV explores all possible path plannings.
%\end{proof}

%Later in experiments, though we implement a novel and optimal algorithm based on integer linear programming (ILP) for finding the path planning for UAV swarm with charging stations in Section \ref{ap:ilp}, it has high computational complexity and can hardly be used in practice for solving large scale UAV-swarm path planning problems.

This proposition tells that our proposed Algorithm~\ref{alg:greedy-charging} has desirable performance bound even in the worst case, while enjoying the low computational complexity. Later in Section~\ref{sec:experiments} we will use extensive simulations in Fig.~\ref{fig-experiment} to show that the performance of our algorithm is near-optimal.

%mention muti-UAV DP, story.
%Mention. System model, expected demand time variance, demand release.
\section{Simulation  Experiments}\label{sec:experiments}

In this section, we implement our proposed  algorithms and conduct extensive simulations to evaluate their performances.
%as compared to the optimum where we implement the integer linear programming algorithm to obtain the optimal routing of UAV-swarm.
%
%\myparagraph{Experiment Setting.}
The data used in simulations are generated randomly to mimic the practical UAV service provisioning.
%and we sample $100$ instances to obtain average performances for each setting.
%we generate $60$
%Specifically, the demands are generated randomly over time span up to $50$ minutes over up to $30$ locations. For each demand, we first pick random values for each demand's duration in the discrete range of $[1,20]$ minutes and set random values for demand release times in the discrete range of $[1,50]$ minutes. 
Specifically, the demands are generated randomly over time span up to $40$ minutes. We first pick random values for each demand's duration in the discrete range of $[1,20]$ minutes and set random values for demand release times and deadlines in the discrete range of $[1,40]$ minutes. 
The UAV's service time for each demand is $2$ minutes. 
For user locations and charging stations, we generate integer value of 2D ground coordinates with uniform distribution in range $[0,10]$ km for both $x$-axis and $y$-axis, and  adopt \emph{Manhattan Distance} to measure their pairwise distance for the UAVs to fly through.
Each UAV's battery capacity is set to be $30$ units of energy, the energy consumption rates for hovering and flying are set to be $2$ and $3$ units per minute, respectively.
We consider $3$ charging stations and it takes $3$ minutes to fast charge or swap the battery.
The above setting applies to all the following experiments unless separately described.

\iffalse
\begin{figure*}[!h]
\subfloat[Cooperative trajectories at optimum with charging]{
\begin{minipage}[t]{0.45\linewidth}
\centering
\includegraphics[height = 60mm]
{images/figure-greedy-energy.pdf}
\label{Fig.trajA.energy}
\end{minipage}
}
\subfloat[Cooperative trajectories returned by our algorithm with charging]{
\begin{minipage}[t]{0.45\linewidth}
\centering
\includegraphics[height = 60mm]
{images/figure-opt-energy.pdf}
\label{Fig.trajB.energy}
\end{minipage}
}
\caption{
Three UAVs cooperative routing over the 2D ground plane including $10$ location nodes with $30$ demands (marked by time interval $[r_j, d_j]$ for demand $j \in J_s$ near each location node $s$). The optimal solution and Algorithm~\ref{alg:greedy}'s solution are shown in Figure~\protect\subref{Fig.trajA.energy} and \protect\subref{Fig.trajB.energy}, respectively. Different UAVs' trajectories are marked in different colors (black, blue, green), and the time interval along each trajectory tells the flight time interval of each UAV. The location nodes with double circles are the initial locations of UAVs.
}
\label{Fig.traj.energy}
\end{figure*}
\fi

\subsection{Performance evaluation of UAV-swarm algorithms without battery charging}

%without charging

%with/without charging
%example. then , data.
%similar pattern.
In this experiment, we present the results without battery charging.
Recall that our Algorithm~\ref{alg:one-general} already returns the optimal solution for a single-UAV case in Section~\ref{sec:single-drone}, and Algorithm~\ref{alg:greedy} extends Algorithm~\ref{alg:one-general} for $|K|$-many cooperative UAVs in Section~\ref{sec:multiple-drone}. We aim to verify the performance of Algorithm~\ref{alg:greedy}, and we implemented a conventional baseline approach where UAV locations are partitioned into clusters and one cluster is assigned to one UAV, as discussed in Section \ref{subsec:benchmark}. 
Specifically, we marge a pair of clusters/locations with the minimum marginal distance into one cluster until a number of $|K|$ clusters achieved \cite{grygorash2006minimum}. For each single UAV, we still apply our optimal Algorithm~\ref{alg:one-general} to obtain the path planning.

%We aim to verify the performance ratio of Algorithm~\ref{alg:greedy} versus the optimum. 
We first verify the performance of Algorithm~\ref{alg:greedy} versus the optimum, here we use average performance ratio metric \cite{stefas2020approximation},
where performance ratio is the ratio of the number of serviced demands by Algorithm~\ref{alg:greedy} to that of optimum.

%and present the performance ratio in Fig.~\ref{fig-experiment}, 
%The performance ratio is averaged by running 100 samples for each data point.
%obtained by ILP in Section~\ref{ap:ilp} in much more running time.
%
%\noindent
%
%
\begin{figure}[!ht]%
\centerline{
\includegraphics[keepaspectratio,width = 0.98\columnwidth]{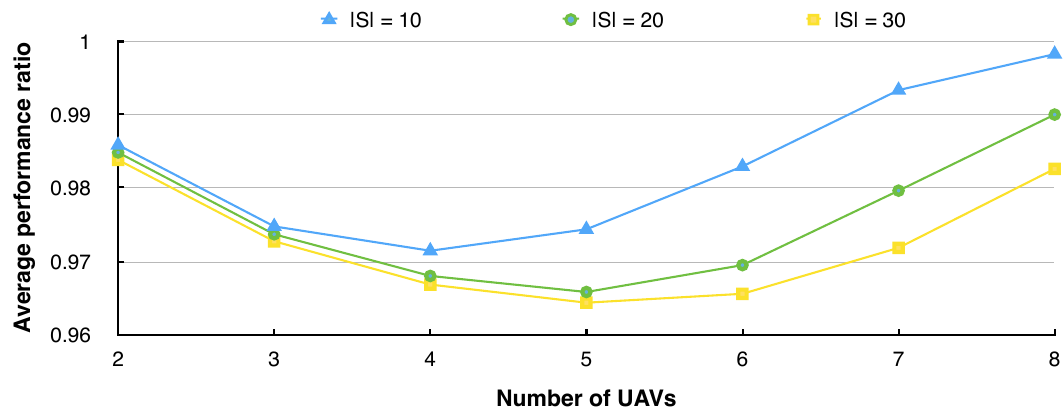}
}
\caption{Average performance ratios achieved by our Fast Iterative UAV-swarm Routing Algorithm versus the optimum under different values of the UAV number $|K|$ and the location number $|S|$.}
\label{fig-experiment}
\end{figure}
%
%
%and with small demand service time, i.e., $q \rightarrow 0$.
%We first present the cooperative UAV Swarm under Algorithm~\ref{alg:greedy} and the optimum for an example with $10$ locations $30$ demands in Fig.~\ref{Fig.traj}\protect\subref{Fig.greedy} and Fig.~\ref{Fig.traj}\protect\subref{Fig.opt}.
%There are some mild differences between these two solutions. In total, the optimum services $26$ demands and our algorithm services $24$ demands with performance ratio $24/26 > 90\%$.
%
%We study how the relative performance of Algorithm~\ref{alg:greedy} versus the optimum changes with the number of UAVs $|K|$ in Fig.~\ref{fig-experiment}.
In Fig.~\ref{fig-experiment}, as the performance of the algorithm is measured in average sense (average over 100 random inputs), it is expected to be greater than the approximation ratio in the worst case analysis as in Proposition \ref{lmm-greedy-cover}.
%In experiments, we vary the value of $|K|$ from 2 to 8, and sample $100$ instances for each value. Besides $|K|$, we vary the number of locations $|S|$ from $10$ to $20$ and $30$.
%The summarized performance results are shown in Fig.~\ref{fig-experiment}.
We can see from Fig.~\ref{fig-experiment} that the average performance ratio of Algorithm~\ref{alg:greedy} is overall high (above $96\%$), and it is very close to 1 (i.e., optimum) when the number of UAVs is small or large enough. When $|K|$ is small, UAVs are sparsely distributed among many user locations and the "separated" UAVs cover very different user locations at optimum. 
Thus, our iterative Algorithm~\ref{alg:greedy} which routes UAVs one by one in a greedy manner performs very close to the optimum. When $|K|$ is large, most user locations are covered by enough UAVs and our algorithm missing very few performs very close to the optimum as well. Furthermore, as the location number $|S|$ increases, the performance gap between the optimum and our algorithm increases as there are more joint routing possibilities among UAVs to consider.

\iffalse
\begin{figure}[!t]%
\centering
\begin{minipage}[t]{0.47\columnwidth}
\centerline{
\includegraphics[width = 0.97\columnwidth]{images/figure_data.pdf}
}
\caption{Performance ratios achieved by our Fast Iterative UAV-swarm Routing Algorithm versus the optimum under different values of the UAV number $|K|$ and the location number $|S|$.}
\label{fig-experiment}
\end{minipage}
\centering
\begin{minipage}[t]{0.47\columnwidth}
\centerline{
\includegraphics[width = 0.97\columnwidth]{images/figure_data_noise.pdf}
}
\caption{Performance ratio achieved by Algorithm~\ref{alg:greedy-charging} with expected demand arrivals versus the random demand arrivals under Gaussian distribution.}
\label{fig-noise-energy}
\end{minipage}
\end{figure}
\fi

\iffalse
%both figures are removed
\subfloat[Trajectories returned by Algorithm~\ref{alg:greedy}]{
\begin{minipage}[t]{0.235\columnwidth}
\centering
\includegraphics[keepaspectratio, width = 1.05\columnwidth]
{images/figure-greedy.pdf}
\label{Fig.greedy}
\end{minipage}
}
\subfloat[Trajectories at optimum]{
\begin{minipage}[t]{0.235\columnwidth}
\centering
\includegraphics[keepaspectratio, width = 1.05\columnwidth]
{images/figure-opt.pdf}
\label{Fig.opt}
\end{minipage}
}
\fi

\begin{figure}[!ht]%
\centerline{
\includegraphics[keepaspectratio,width = 0.98\columnwidth]{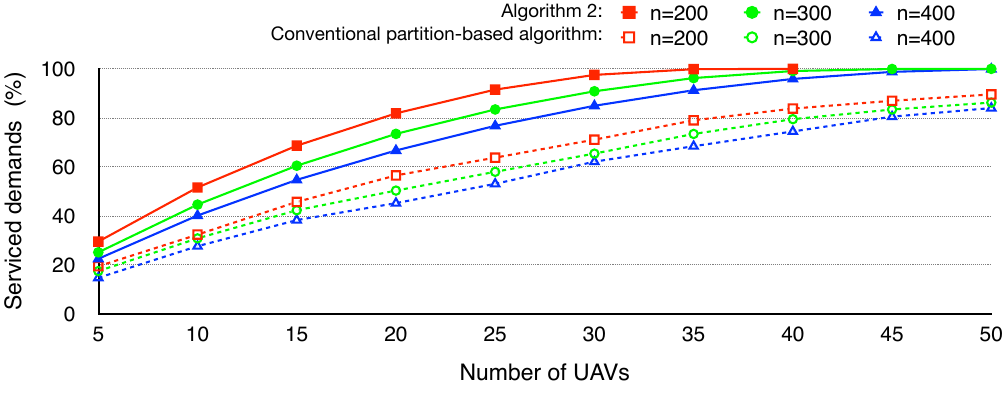}
}
\caption{
Percentage of serviced demands versus the number of UAVs
for $|S|=100$ locations by Algorithm~\ref{alg:greedy} (solid lines) and by the conventional partitioned-based algorithm in Section \ref{subsec:benchmark} (dashed lines), respectively.
}
\label{fig-serviced-percentage}
%n=400,K=100,P=50,beta=5,q=1,[1,50); {'xyrange': 30, 'weight': False, 'prop': [0.0, 2]}
\end{figure}

We then verify the performance of Algorithm~\ref{alg:greedy} with the baseline approach mentioned above for large scale input of $|S|=100$ locations (which is hard to compute the optimum). The coordinates of user locations are randomly selected from $[0,30]$ km to support up to $n=400$ demands release.
Fig.~\ref{fig-serviced-percentage} shows how the percentage of serviced demands changes with the increase number of UAVs.
%We can see from the figure that Algorithm~\ref{alg:greedy} significantly outperforms the baseline approach, especially for large number of UAVs.
When the UAV number becomes large and sufficient to service all the demands, the serviced demands rate tends to converge. Besides, Algorithm~\ref{alg:greedy} significantly outperforms the conventional approach for various setting of UAV number $|K|$.

%\subsection{Impact of battery charging on UAV path planning}
\subsection{
Performance evaluation of UAV-swarm algorithms with battery charging} \label{subsec:exp-2}

\begin{figure}[t!]
\centering
\subfloat[UAVs' Trajectories returned by Algorithm~\ref{alg:greedy-charging}  with charging]{
\begin{minipage}[t]{0.98\columnwidth}
\centering
\includegraphics[keepaspectratio, width = 0.98\columnwidth]
{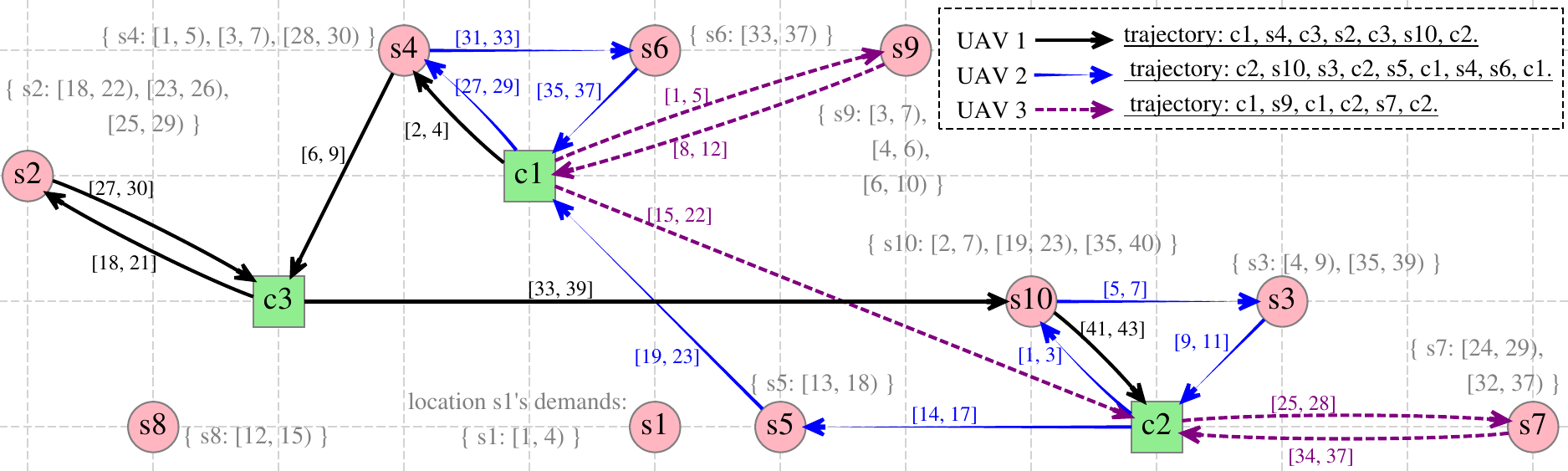}
\label{Fig.greedy.energy}
\end{minipage}
}
\newline
\centering
\subfloat[UAVs' Trajectories at optimum with charging]{
\begin{minipage}[t]{0.98\columnwidth}
\centering
\includegraphics[keepaspectratio, width = 0.98\columnwidth]
{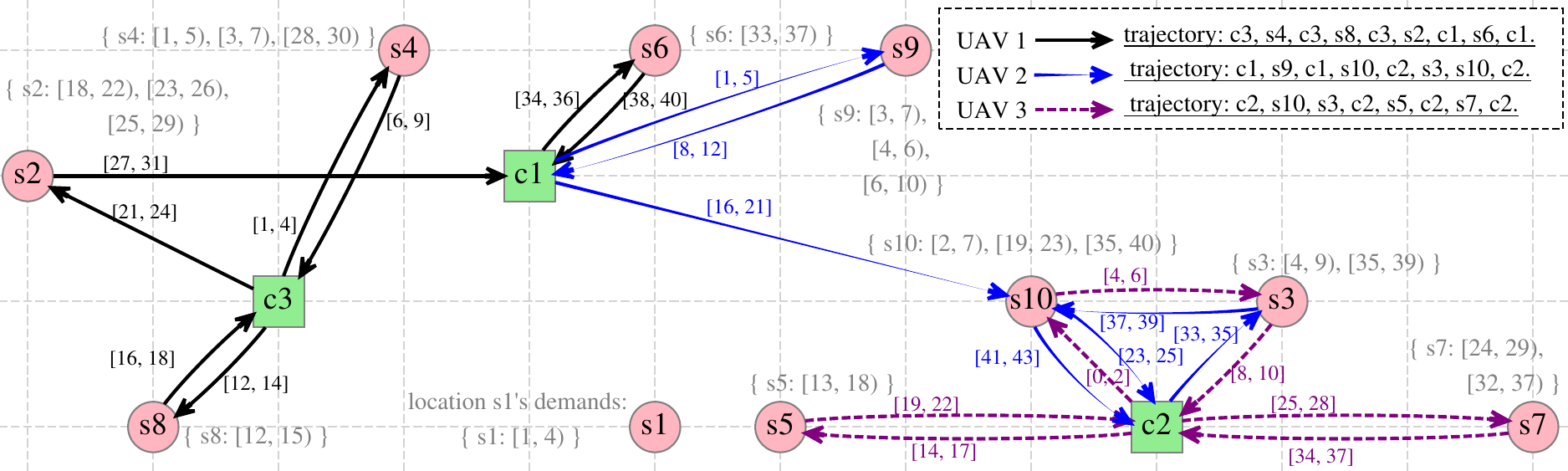}
\label{Fig.opt.energy}
\end{minipage}
}
\caption{
Three UAVs' cooperative path planning over 2D ground plane across $10$ locations $s_1$ - $s_{10}$ and battery charging stations $c_1$ - $c_{3}$, where the circle nodes indicate user locations and square nodes indicate charging stations.
%Fig.~\protect\subref{Fig.greedy} and \protect\subref{Fig.opt} correspond to the problem without charging and near-zero demand service time, and they represent the routing solutions returned by Algorithm~\ref{alg:greedy} and by the optimal algorithm, respectively.
Fig.~\ref{Fig.traj}\protect\subref{Fig.greedy.energy} and \protect\subref{Fig.opt.energy} show each UAV's trajectory (over different nodes) returned by Algorithm~\ref{alg:greedy-charging} and by the optimal algorithm, respectively.
%in Appendix~\ref{ap:ilp}
%(n,S,K) = (20, 10, 3)
%(minT, maxT) = (1, 40)
%(beta) = (10)
%(q,C,B) = (2, 3, 30)
%P = [2,3]
%The battery capacity is $30$,  service time $q=2$, charging time $t_c=2$, alg=15 vs. opt=17
%line shapes with
The three UAVs' trajectories are marked in different colors (black, purple, blue).
%, and the time interval along each trajectory tells the flight time interval of each UAV
Near each location node $s$, the demand $j \in J_s$ is depicted as time interval $[r_j, d_j)$ 
with arrivel time $r_j$ and deadline $d_j$, and for example location $s_3$ faces two demand windows $[4, 9)$ and $[35, 39)$ minutes.
Each directed edge connecting two nodes corresponds a routing decision for the corresponding UAV, where the label $[t_1,t_2]$ for each edge indicates that the UAV departs from one node at time $t_1$ and arrives at the other node at time $t_2$.
%The legends in the upper-right corner contain the sequence of locations that each UAV has visited in the path planning.
%Three UAVs cooperative routing over the 2D ground plane including $10$ location nodes with $30$ demands (marked by time interval $[r_j, d_j]$ for demand $j \in J_s$ near each location node $s$). The optimal solution and Algorithm~\ref{alg:greedy}'s solution are shown in Figure~\protect\subref{Fig.opt} and \protect\subref{Fig.greedy}, respectively.
%The location nodes with double circles are the initial locations of UAVs.
}
\label{Fig.traj}
\end{figure}

%in Appendix~\ref{ap:ilp}
In this subsection, we examine our low-complexity Algorithm~\ref{alg:greedy-charging}'s average performance and compare it to our optimal algorithm based on ILP as detailed in Appendix \ref{ap:ilp}.
%
%In previous section, we show the average performance ratios archived by our algorithms, which confirms our analytical theoretical bounds. Here, we study how battery charging affects UAV-swarm cooperation in the path planning. From extensively tested examples, we found that  Algorithm~\ref{alg:greedy-charging} can service as many demands as optimum with much high probability.
%in Appendix~\ref{ap:ilp}
Here, we present one typical example of three UAVs' cooperative trajectories with three charging stations, where the solutions returned by our Algorithm~\ref{alg:greedy-charging} and our optimal ILP algorithm are depicted in
Fig.~\ref{Fig.traj}\protect\subref{Fig.greedy.energy} and Fig.~\ref{Fig.traj}\protect\subref{Fig.opt.energy}, respectively.
%in the 2D ground plane
With $20$ user demands released over $10$ locations, the optimum services $17$ demands and our Algorithm~\ref{alg:greedy-charging} services $16$ demands with average performance ratio $94\%$ 
as compared to the optimum. Though the trajectories in these two sub-figures look somewhat different, in both the optimum and Algorithm~\ref{alg:greedy-charging}'s solution, we observe smart UAVs' cooperation by overlapping their trajectories in certain user locations.

%which is consistent with the results in
%We first observe that the UAV could travel directly from one charging station to another charging station.

Next, we study how battery charging affects UAV-swarm cooperation in the path planning. 
As shown in Fig.~\ref{Fig.traj}\protect\subref{Fig.greedy.energy}, UAV 3 flies from charging station $c_1$ to another charging station $c_2$, before reaching location $s_7$ to service the demands there.
%UAV 1 flies from charging station $c_1$ to another charging station $c_3$, before reaching location $s_2$ to service the demands there.
One may wonder why the UAV needs to charge again at charging station $c_2$ during the flight.
%Intuitively, charging twice compensates the energy consumption during the flight from $c_1$ to $s_2$, especially when the stopping point $c_3$ is close to $s_2$.  
Intuitively, charging twice compensates the energy consumption during the flight from $c_1$ to $s_7$, especially when the stopping point $c_2$ is close to target user location $s_7$.  
Moreover, even with the battery charging option, our optimal algorithm using ILP chooses not to service location $s_1$ which is far from the other nodes, and similarly our sub-optimal Algorithm~\ref{alg:greedy-charging} chooses not to service location $s_8$ at the corner.

%
%Overall, Algorithm~\ref{alg:greedy-charging} takes about 3.2 seconds to compute the path plannings of servicing 60 demands, compared to that of 437 seconds for the ILP algorithm, on the compute platform with Intel Core i5-3570 CPU.
%这里简单说下运算时间的差距有多大，可以按之前结果自己估计一下就好了，没必要重新做实验再算。
%While, the performance ratio of Algorithm~\ref{alg:greedy-charging} is much better than the theoretical bound in worst case, and it is close to 1 in practice (i.e., optimum). By this, our proposed Algorithm~\ref{alg:greedy-charging} is more suitable for solving the UAV-swarm path planning problem with battery charging in practice.

\subsection{Algorithm robustness to noisy demand information}

%demands are known.
%full knowledge each hotspot
One may wonder the performances of our proposed algorithms if the UAVs do not have precise information about the demands to service at each location.
This is possible in practice if the demand forecasting has some error. We choose Algorithm~\ref{alg:greedy-charging} as it fits the most general scenario with battery charging stations and conduct experiments to examine its robustness to noisy demand information.
We refine our Algorithm~\ref{alg:greedy-charging} 
to use the expected demand information such as mean arrival time as input to obtain the cooperative path planning solution, while the actual arrival of each demand may deviate from the mean with some noisy Gaussian distribution in the time domain.
%some particular variance.
%We show how this variance affects the performance of the algorithm.
We show how this noisy demand distribution affects the performance of the algorithm, by counting the finally serviced demands.

\begin{figure}[!ht]%
\centerline{
\includegraphics[keepaspectratio, width = 0.98\columnwidth]{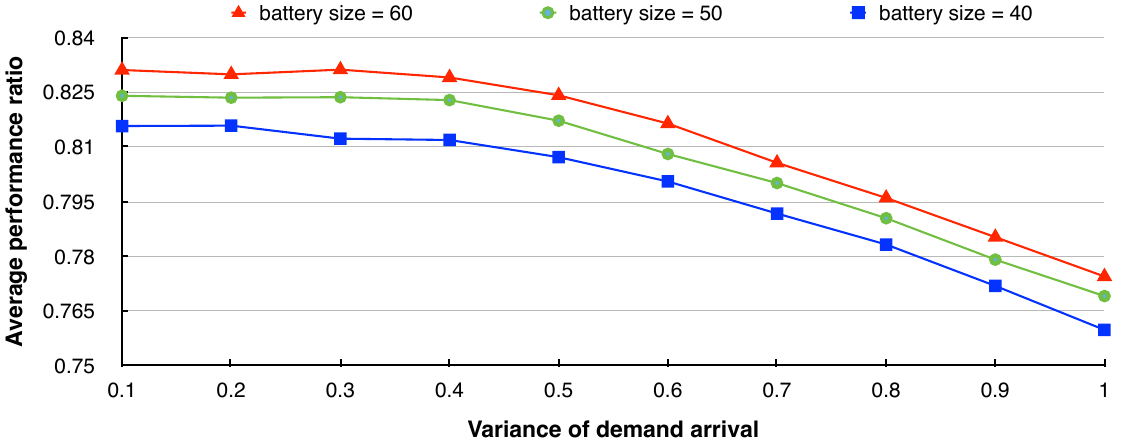}
}
\caption{Average performance ratio between Algorithm~\ref{alg:greedy-charging} and the offline optimum (with precise information) versus the variance of the demand arrival information. 
}
\label{fig-noise-energy}
\end{figure}

%n=60, B = 40.
%The result is shown in  
Fig.~\ref{fig-noise-energy} presents the average performance ratio of Algorithm~\ref{alg:greedy-charging}, as compared to the offline optimum which knows precisely all the demand information. 
Naturally, more demands are missed given a larger variance of the demand information unknown to the UAV-swarm. We can see that as compared to the offline optimum, Algorithm~\ref{alg:greedy-charging} performs well given reasonable variance of noisy demand information, with average performance ratio above $75\%$ for various settings of energy capacity. Though sometimes a UAV may leave a location early to miss a delayed demand, due to the trajectory overlaps among the cooperative UAVs, this demand can still be serviced by another UAV flying to the same location later.

\iffalse
\begin{figure}[t]%[!h]
%\subfloat[Cooperative trajectories returned by Algorithm~\ref{alg:greedy}]{
\centering
\begin{minipage}[t]{0.7\linewidth}
\centering
\includegraphics[height = 60mm]
{images/figure_trajB.pdf}
\label{Fig.trajB}
\end{minipage}
%}
\caption[12pt]{
Trajectories returned by Algorithm~\ref{alg:greedy} where three UAVs cooperative routing over the 2D ground plane including $10$ location nodes with $30$ demands (marked by time interval $[r_j, d_j]$ for demand $j \in J_s$ near each location node $s$).
%The solution returned by Algorithm~\ref{alg:greedy} is shown in Figure~\protect\subref{Fig.trajB}.
Different UAVs' trajectories are marked in different colors (black, blue, green), and the time interval along each trajectory tells the flight time interval of each UAV. The location nodes with double circles are the initial locations of UAVs.
}
\label{Fig.traj}
\end{figure}
\fi

%n=60,K=30,P=8,beta=3,[1,30)

%On the other hand, when there are more locations, the performance of the algorithm becomes worse.
%This maybe the reason that with more locations,

\section{Conclusion and Future Directions}
In this paper, we study the UAV-swarm routing problem for servicing many spatial locations with dynamic user arrivals and waiting deadlines in the time horizon, aiming to maximize the total number of successfully serviced user demands. For a single UAV's routing problem, we propose an optimal algorithm of fast computation time. When a large number of UAVs are coordinating, the problem becomes intractable and we prove that the performance of partitioning UAVs to service separate location clusters can be arbitrarily poor. Alternatively, we present an iterative cooperative routing algorithm with constant factor approximation ratio in the worst case.
Moreover, to further relax UAVs' energy capacity limit for sustainable service provisioning, we allow UAVs to travel to charging stations in the mean time and thus jointly design UAVs' path planning over users' locations and charging stations.
Based on the DAG of the UAV-state transition diagram, we successfully transform the problem to an integer linear program to solve it optimally.
%optimally solve the problem.
To achieve low computational complexity, we further propose an iterative algorithm with constant approximation ratio with energy charging.
Finally, we validate the theoretical results by extensive simulations.
%

%The major contribution of our work is to provide an optimal framework for mobile service of UAVs across space and time.
%%although there are still some assumptions in our work that seems ideal.
%The difference of this problem with existing typical vehicle routing problem is that the UAVs in our model need repeatedly visit some locations. And to exploit such properties, we designed fast algorithms with low time complexity, which only depends on the number of demands, regardless of the scale of time domain.
%Moreover, we have shown that traditional ideas of reducing the complexity of the problem by partitioning UAVs can leads to poor performance in the worst case. Our proposed algorithm, instead, jointly considers the cooperation of UAVs with  provable constant approximation performance guarantee.

In future work, we may consider the scenario with completely unknown users' demands where the demands may be released in an online version, and we aim to design online algorithms with provable competitive ratio.

% or


%% use section* for acknowledgment
%\ifCLASSOPTIONcompsoc
%  % The Computer Society usually uses the plural form
%\else
%  % regular IEEE prefers the singular form
%\fi

\ifCLASSOPTIONcaptionsoff
  \newpage
\fi



%\bibliographystyle{IEEEtran}

\bibliographystyle{IEEEtran}


\newpage

%\clearpage
\appendices

\section{Optimal Integer Linear Programming (ILP) Algorithm with energy charging} \label{ap:ilp}

%\noindent
%Due to page limit, the description of our ILP approach to compute the optimal UAV-swarm with energy charging can be found in \cite{wang:2020}.

%\iffalse
When we relax UAVs' energy capacity limit for sustainable service provisioning, UAVs are allowed to travel to charging stations in the mean time and thus jointly design UAVs' path planning over users' locations and charging stations. Multiple path planning should take into account of UAVs' temporal-spatial information of charging and its design becomes more challenging. Recall that the dimensionality-reduced dynamic programming algorithm in previous sections finds the optimal path planning of a single UAV by extending the decision state at each critical time stamp until reaching the decision state where no more demands can be serviced by the UAV.

%To tackle the challenge, we focus on all the refined feasible path plannings for a single UAV.
%As the UAVs are identical, any feasible path planning for one UAV is also feasible for other UAVs.
%Moreover, we observe that each feasible path planning for a single UAV can be represented by a path on a directed acyclic graph.
%%When merging the path plannings for all UAVs, we focus on reducing the common serviced demands to achieve the best collaboration between UAVs.

Thus, we find a novel way to construct a directed acyclic graph (DAG) to represent the whole state transition diagram of the algorithm.
That is to say, each directed path on the DAG corresponds to a feasible path planning for a single UAV.
Therefore, solving the problem of $|K|$ UAVs is simplified to select $|K|$ paths on the DAG.
We then apply integer linear programming approach to select such multiple paths on the DAG globally to find the optimal solution.
Although the DAG contains all feasible path plannings, the number of nodes in DAG (i.e., the number of decision states) can be effectively reduced by our previous analysis of the dimensionality-reduced dynamic programming approach in Section~\ref{sec:single-drone}.

To be specific, We first describe the construction of the DAG $G = (V,E)$  representing the state transition diagram, in which any path from the root node to the leaf node indicates a feasible path planning of a single UAV.
%We exploit this DAG graph to solve the UAV-swarm via integer linear programming approach as follows.
%We use a directed acyclic graph $G = (V,E)$ to store all feasible path plannings of a single UAV, in which any path from the root node to the leaf node indicates a feasible path planning.
%with computed variable $g(Q,i,s,t) = B_t$
Specifically, each node $v \in V$ corresponds to a decision state $v = \pe{Q, B_{t}, i, s, t}$ as described in Section~\ref{subsec:energy-state} and each directed edge $e = (v,v') \in E$ connecting two nodes $v,v' \in V$ indicates a routing action that transits the decision state $v = \pe{Q, B_{t}, i, s, t}$ into the new decision state $v' = \pe{Q', B_{t'}, i', s', t'}$.
Such a graph can be generated based on our proposed dimensionality-reduced dynamic programming approach in Section~\ref{subsec:energy-state}.
Recall that the decision state $v$ will be disregarded if the battery energy $B_{t}$ is not sufficient enough for the UAV to reach the nearest charging station after time $t$, i.e., $B_{t} < p_1 \cdot \min_{c \in C} a(s,c)$.
In the following of this section, notation $e$ is only used to refer to a directed edge in the graph.

%
%\myparagraph{Integer Linear Programming Approach}
%Given the directed acyclic graph $G = (V,E)$, we aim to find the optimal solutions for multiple drones.
%As mentioned above, each node $v = (\bd{u},s,t)$ is defined by three parameters, pending tasks defined by $\bd{u}$, location $s \in S$ and time $t \in \mathcal{T}$.
Now we describe the integer linear programming approach to find the optimal solution for multiple UAVs, where we regard the solution of each UAV as an independent path planning, i.e., the same demand might be serviced by more than one UAV.
%
%For each location $s \in S$, for each node $v \in V$, we write $s(v)$ indicates the location that is contained in node $v$.
We denote ${root}$ as the entrance of the dimensionality-reduced dynamic programming algorithm, which is also the only root node of the graph.
For each node $v \in V$, let $\text{IN}(v)$ and $\text{OUT}(v)$ be the set of incoming edges and outgoing edges connected to node $v$, respectively.
%For each user demand $j \in J$, let $V(j) \subseteq V$ be the set containing any node $v$ in which demand $j$ is serviced according to the decision state $v = \pe{Q, B_{t}, i, s, t}$ up to time $t$, i.e. $j \in J(v)$, and let $E(j) = \bigcup_{v \in V(j)} \text{IN}(v) \cup \text{OUT}(v)$ be the set of edges that contains any node from $V(j)$.
Recall that each routing action will result in at least one new demand to be serviced, hence each transition (i.e., edge in the graph) corresponds to the additionally serviced demands by that routing action.
Then, for each user demand $j \in J$, we denote $E(j) \subset E$ as the set containing any edge $e = (v,v')$ such that demand $j$ is one of the additional demands serviced by the UAV at location $s'$ when transiting from decision state $v = \pe{Q, B_{t}, i, s, t}$ to  $v' = \pe{Q', B_{t'}, i', s', t'}$.
%, i.e., one of the demands represented by $i'-i$.
%
For each UAV $k \in K$ and each edge $e = (v,v') \in E$ from the graph, we create a binary decision variable $x_{k,e}$, where $x_{k,e} = 1$ indicates that the path that corresponds to the solution of UAV $k$ contains the edge $e$ in the graph, i.e., the UAV indeed encounters decision state $v$ and transits to decision state $v'$.
%the decision state $v'$ is indeed transited from state $v$.
For each UAV $k \in K$ and each demand $j \in J$, we create a binary decision variable $b_{k,j}$ where $b_{k,j} = 1$ indicates that demand $j$ is serviced by UAV $k$.
Moreover, we create binary variable $o_j$ for each demand $j \in J$ where $o_j = 1$ indicates that demand $j$ is serviced by at least one UAV.
The integer linear programming is formulated as follows.

\begin{align}
  \max   \sum_{j \in J} o_j \label{eq:obj} &~\\
  \sum_{e \in \text{OUT}(v)} x_{k,e} - \sum_{e \in \text{IN}(v)} x_{k,e}  \le 0, &~ \forall k \in K, v \in V\setminus\se{\text{root}} \label{eq:1}\\
  \sum_{e \in \text{OUT}(v)} x_{k,e} \le 1,  &~  \forall k \in K, v = {\text{root}} \label{eq:2}\\
  b_{k,j} - \sum_{ e \in  E(j)} x_{k,e} \le 0, &~ \forall k \in K,  \forall j \in J \label{eq:3}\\
  o_{j} - \sum_{k \in K} b_{k,j} \le 0, &~  \forall j \in J \label{eq:4} 
  \\
  x_{k,e}, b_{k,j}, o_j  \in \se{0,1}, &~ \forall k \in K, e \in E, j \in J \label{eq:5}
  %\\
  %\sum_{k \in K} \sum_{e \in \text{OUT}(v)} x_{k,e}   \le 1, &~ \forall v \in V\setminus\se{\text{root}} \label{eq:5}
\end{align}
In the above formulation,
Equation~\eqref{eq:obj} defines the objective function, which indicates the total number of serviced demands.
Equation~\eqref{eq:1} and \eqref{eq:2} guarantees that each UAV corresponds to a path starting with the \textit{root} node and ending with any node in the graph. 
Equation~\eqref{eq:3} ensures that if demand $j$ is serviced by UAV $k$, at least one edge from $E(j)$ is selected in the path.
Equation~\eqref{eq:4} ensures that demand $j$ is counted as serviced only if at least one UAV can service the demand.
%Equation~\eqref{eq:5} is an optional constraint which means that two UAVs can not share the same node, which is correct since each node $v = (Q,s,t)$ indicates that the UAV is at location $s$ at time $t$. If two UAVs share the same node, it means the two UAVs are at the same location at the same time, which should definitely be avoid.

After solving the above ILP, the path of UAV $k$ can be directly constructed by including the edge $e$ as long as $x_{k,e} = 1$.
Based on that, the path planning of the UAV $k$ can be constructed according to the decision state of the nodes in the path. In Section~\ref{sec:experiments}'s experiment, we use Gurobi as the ILP solver.
%\fi


%\vfill


\end{document}